\documentclass[11pt, oneside,reqno]{amsart}   	
\usepackage[margin=2cm]{geometry}                		
\usepackage{amsaddr}
\usepackage{graphicx}
\usepackage{bm}
\usepackage{import}
\usepackage{amsmath} 
\usepackage{amsthm}
\usepackage{amsfonts}
\usepackage{amssymb}
\usepackage{comment}
\usepackage{braket}
\usepackage{mdframed}
\usepackage{algpseudocode}
\usepackage{algorithm}
\usepackage{float}
\usepackage{tikz}
\usepackage[dvipsnames]{xcolor}
\usepackage{subcaption}
\usepackage{mathrsfs}

\usepackage{cite}
\usepackage[T1]{fontenc}

\usepackage{framed}
\definecolor{shadecolor}{gray}{0.95}
\usepackage{hyperref}
\usepackage{xfrac}

 \newcommand{\eps}{\varepsilon}

  \hypersetup{
    colorlinks=true,
    linkcolor=blue,
    citecolor=blue,
    filecolor=black,
    urlcolor=blue,
}
\usepackage[capitalize]{cleveref}

\newtheorem{result}{Result}

\begin{document}
\title[MPS skeletons in Onsager-integrable chains]{Matrix-product state skeletons in Onsager-integrable quantum chains.}
\author{Imogen Camp}
\address{Rudolf Peierls Centre for Theoretical Physics, University of Oxford, Oxford, UK}
\email{imogen.camp@physics.ox.ac.uk}
\author{Nick G. Jones}
\address{St John’s College and Mathematical Institute, University of Oxford, Oxford, UK}
\email{nick.jones@maths.ox.ac.uk}
\thanks{The published version of this article is J. Stat. Phys. 193, 73 (2026); \href{https://doi.org/10.1007/s10955-026-03637-8}{doi.org/10.1007/s10955-026-03637-8}.}

\begin{abstract}
Matrix-product state (MPS) skeletons are connected networks of Hamiltonians with exact MPS ground states that underlie a phase diagram. Such skeletons have previously been found in classes of free-fermion models. For the translation-invariant BDI and AIII free-fermion classes, it has been shown that the underlying skeleton is dense, giving an analytic approach to MPS approximation of ground states anywhere in the class. In this paper, we partially expose the skeleton in certain interacting spin chains: the $N$-state Onsager-integrable chiral clock families. We construct MPS that form a dense MPS skeleton in the gapped regions surrounding a sequence of fixed-point Hamiltonians (the generators of the Onsager algebra). Outside these gapped regions, these MPS remain eigenstates, but no longer give the many-body ground state. Rather, they are ground states in particular sectors of the spectrum. Our methods also allow us to find further MPS eigenstates; these correspond to low-lying excited states within the aforementioned gapped regions. This set of MPS excited states goes beyond the previous analysis of ground states on the $N=2$ free-fermion MPS skeleton. As an application of our results, we find a closed form for the disorder parameter in a family of interacting models. Finally, we remark that many of our results use only the Onsager algebra and are not specific to the chiral clock model representation.
\end{abstract}
\maketitle
\setcounter{tocdepth}{1}
\tableofcontents\clearpage
\section{\label{sec:intro}Introduction}
In recent decades, significant advances have been made in understanding the ground state physics of one-dimensional quantum many-body systems. While integrability provides an analytical probe of phenomena in special families of Hamiltonians \cite{Korepin97,Baxter16}, numerical methods related to matrix-product state (MPS) approximations allow us to understand features of the general interacting case \cite{White92,Schollwock11}. When the many-body Hamiltonian has a spectral gap, the entanglement area law guarantees an MPS approximation to the ground state \cite{Hastings07,Cirac21}. Beyond approximations, MPS are also a valuable analytical tool. MPS-solvable Hamiltonians have exact MPS ground states, in the sense that the ground state can be represented with a finite bond dimension in the thermodynamic limit. Such models, including the AKLT chain \cite{Affleck87}, have led to many insights \cite{Cirac21}. For example, MPS approaches have led to a deeper understanding of symmetry fractionalisation, order parameters, and the classification of symmetry-protected topological (SPT) phases in one dimension \cite{Pollmann10,Chen11,Schuch11,Turner11}.

It is interesting to consider the connection between integrable models and Hamiltonians that admit exact MPS ground states: neither property implies the other, yet both facilitate analytic insights into quantum many-body systems. Simple examples of MPS-solvable Hamiltonian families are the one-parameter ``disorder line'' of product states in the XY model \cite{Barouch71,Kurmann82,Muller85,Chung01,Franchini07}, and its MPS-solvable Kramers-Wannier dual \cite{Wolf06,Smith22}. 
Refs. \cite{Jones21,Jones23} investigated this question more broadly in families of free-fermion chains (the translation-invariant BDI and AIII classes of the tenfold way \cite{Altland97}). 
Despite the simple exact-solvability of free fermion models \cite{Lieb61}, their ground states are generically \emph{not} exact matrix-product states, as their entanglement spectra have infinite rank \cite{Franchini11,Jones21}. However, taking the BDI class for definiteness, we can precisely characterise the subset of this family with exact MPS ground states. Each model in the BDI class corresponds to a Laurent polynomial $f(z)$, and the model has an exact MPS ground state if and only if $f(z)= z^p g(z)^2 h(z)$ for some integer $p$, polynomial $g(z)$, and Laurent polynomial $h(z)$, where $h(z)=h(1/z)$. The sub-family satisfying this condition constitutes the MPS skeleton: a structure underlying the entire phase diagram. In contrast to finding an MPS approximation to the ground state of a general gapped Hamiltonian, the ground states along this skeleton can be represented exactly by an MPS with finite bond dimension in the thermodynamic limit. While lying exactly on the skeleton is fine-tuned, it is dense in the following sense: we can find a sequence of models lying on the skeleton with ground states that give an arbitrarily good approximation to the energy density of a generic ground state in the class. Hence, we have an analytic approach for approximating those ground states in the phase diagram that have an infinite-rank entanglement spectrum \cite{Jones21,Jones23}.

It is natural to ask if there are other families of models where the underlying MPS skeleton can be identified. It is of particular interest to go beyond free-fermionic cases. In this work, we construct MPS skeletons underlying interacting families of Onsager-integrable chiral clock models. These are families of integrable models on chains of $N$-state sites, where the Onsager algebra yields local conserved charges \cite{Onsager44,Dolan82,Ahn91}. Taking $N=2$, we recover the BDI class (in its spin chain form). Onsager's 1944 computation of the free energy in the 2D classical Ising model utilised this algebra, and it is thus naturally connected to the 1D quantum transverse-field Ising model and its free-fermion solvability. Nevertheless, other representations of this algebra appear as $N$-state generalisations of the quantum Ising model. For $N>2$, these models are interacting chiral clock models \cite{vonGehlen85, Baxter88, Baxter89, Davies90}. Commutators of the Onsager paramagnet, denoted $A_0$, and the Onsager ferromagnet, denoted $A_1$, generate this algebra. Beyond the superintegrable chiral Potts chain $A_0+\lambda A_1$, the analogue of the transverse-field Ising model, such models remain relatively unexplored. Ref. \cite{Ahn91} examines the Hamiltonian family constructed as a linear combination of the generators, denoted $A_k$, while Refs. \cite{Jones24,Jones25} analyse their SPT physics. Note that in some cases the Onsager generators are considered as a symmetry algebra of some other Hamiltonian, rather than as Hamiltonians themselves \cite{Vernier19,Gioia25,Pace25,Su25}.

As demonstrated in \cite{Davies90}, we can express generators of (general) representations of the Onsager algebra as a direct sum of representations of ${sl}(2)$. In principle, we can use this to diagonalise these models, establishing the \emph{form} of the spectrum \cite{Davies90,Ahn91,Roan11}. In the chiral clock models, these representations are all two-dimensional \cite{McCoy91,VonGehlen01}. Further analysis is required to determine the spectrum from functional equations in the chiral clock case \cite{Baxter89,Albertini89b,McCoy91,Dasmahapatra93,VonGehlen01,Iorgov10}---see also Refs. \cite{AuYang08,AuYang09,Roan11} for more recent work on eigenvectors. These eigenvectors are, roughly speaking, found by filling an interacting Fermi sea \cite{McCoy91,Albertini89,Albertini89b}. While the ground-state phase diagram for the $N=2$ case is well understood \cite{Verresen18,Jones19}, we have a far more limited understanding of the phase boundaries for $N>2$. However, we expect gapless regions comprising a sequence of first-order transitions between different Onsager sectors, as occurs when tuning $A_0 + \lambda A_1$ \cite{Albertini89,Albertini89c, vonGehlen96}. Numerical results from \cite{Jones24} demonstrate this extended gapless region in a two-parameter phase diagram of $A_0+\lambda A_1 + \mu A_2$ (see also \cref{fig:phase-diagrams-d=1}).

For each $N$, we consider the family of Hamiltonians \begin{align}H_A[\{t_m\},N]= \sum_m t_m A_m \qquad t_m\in\mathbb{R}\label{eq:family}\ .\end{align} 
For $N=2$, these spin chains are generalised cluster models \cite{Suzuki71,Keating04,Smacchia11,deGottardi13,Ohta16,Verresen17,Verresen18}
\begin{align}H_A[\{t_m\},2]= - \frac{t_0}{4} \sum_j \sigma^x_j - \frac{t_1}{4}\sum_j \sigma^z_j\sigma^z_{j+1}- \frac{t_{-1}}{4}\sum_j \sigma^y_j\sigma^y_{j+1} +\frac{t_2}{4} \sum_j \sigma^z_{j-1}\sigma^x_j\sigma^z_{j+1} +\dots \ ,\end{align}
where $\sigma_j$ are the usual Pauli operators on site $j$.
For the chiral clock models, we show below that each $A_m$ is a local Hamiltonian with a range of $|m|$ sites (i.e. is supported on $|m|+1$ sites), and we impose that the overall model is finite-range. This means we allow only finitely many non-zero $t_m$. 

Within the family \eqref{eq:family}, we let 
\begin{align}
    \mathcal{S} = \{ H_A[\{t_m\},N] \textrm{~such~that~} H_A \textrm{~connects~to~some~} A_k \textrm{~along~a~gapped~path~within~the~family}  \}; \label{eq:S}
\end{align} i.e., $\mathcal{S}$ is the set of Hamiltonians that are in the gapped regions surrounding each of the Onsager generators $A_k$. The $A_k$ are fixed-point Hamiltonians, with zero-correlation-length ground states \cite{Jones24}. Then, writing $f(z) = \sum_m t_m z^m$, we show that if $f(z) = \pm z^p g(z)^2$, and we are in $\mathcal{S}$, then we can construct an exact MPS ground state except at a measure zero set of points. Thus, we have \emph{an} MPS skeleton for $N>2$, giving a significant generalisation of the $N=2$ results to the interacting case. However, importantly, we note the implication is in one direction, and there may be other Hamiltonians in $\mathcal{S}$ with exact MPS ground states (this is certainly the case for $N=2$, as we have excluded cases with $h(z)\neq \text{const}.$). We also show that this skeleton is dense within $\mathcal{S}$, giving us an analytic approach to approximating the ground state by MPS in this region. A further result of our analysis is the first excited state at zero momentum for models on the skeleton, along with the lowest-energy eigenstate in each non-zero momentum sector, up to the points where these states become degenerate with others in the same sector.  This is a direct analogue to the appearance of an exact MPS ground state in the zero momentum sector: usually, we can approximate gapped ground states in finite-momentum sectors to be of this form up to small corrections \cite{Haegeman12,Haegeman13}.

As with the spectrum, computing correlations in the ground state of $H_A$ is not straightforward \cite{Baxter06}. Remarkable closed formulae exist for the long-distance behaviour of the order and disorder parameters along the line $A_0+\lambda A_1$ \cite{Albertini89c,Baxter05,Iorgov10}. While the class $H_A$ can be analysed using the Bethe ansatz, MPS-solvability gives an alternative approach to computing order parameters in the more general family $H_A$, and we demonstrate this for $H_A = A_0+2 a A_1 + a^2 A_2$.

Many of our results rely only on the algebraic relations, and hence apply in principle to other Onsager-integrable Hamiltonians (subject to some additional assumptions given in \cref{remark}). Of course, any unitary transformation applied to the generators $A_k$ will preserve the algebra and the interpretation of the $A_k$ as Hamiltonians. Our results can be simply applied to such cases. For example, in Ref.~\cite{Ahn91}, the authors generate other families of Hamiltonians by combining unitary pivots \cite{Tantivasadakarn23,Jones24} with Kramers-Wannier duality \cite{Kramers41,Aasen20}. It would be interesting to find a model where the $sl(2)$ representations have dimension greater than two \cite{McCoy91}, and to connect to the constructions in Refs. \cite{Fjelstad11,Minami21,Miao22}. For the remainder of the paper we focus our attention on the chiral clock representations.

We structure the paper as follows. In Section \ref{sec:results}, we summarise key properties of our model and state our main results. Section \ref{sec:examples} illustrates these results with some explicit examples. Section \ref{sec:spectrum} gives more details of the diagonalisation of Onsager-integrable Hamiltonians. In Section \ref{sec:analysis}, we prove the results of Section \ref{sec:results}. We conclude in Section \ref{sec:outlook} with a discussion of possible directions for future work.

\section{\label{sec:results}Key definitions and statement of results}
\subsection{Onsager-integrable Hamiltonians} \label{subsec:chiral-clock-intro}

In Ref. \cite{Onsager44}, Onsager solved the two-dimensional classical Ising model by computing the spectrum of the transfer matrix, giving the exact partition function. Of interest to us are the corresponding quantum Hamiltonians on a chain, which can be determined by taking a strongly anisotropic limit of the two-dimensional classical model. His solution utilised what is now called the Onsager algebra: the infinite-dimensional Lie algebra obeyed by the generators $\{A_l, G_m| l,m \in \mathbb{Z}\}$ given by

\begin{align} \label{eqn:onsager-algebra}
    [A_l, A_m] =  G_{l-m} \quad\quad\quad
    [G_l, A_m] = \frac{A_{m+l}-A_{m-l}}{2}\quad\quad\quad
    [G_l, G_m] = 0\ .
\end{align}

\noindent We observe that the mappings $A_m \mapsto A_{m+1}, G_l\mapsto G_{l}$ and $A_m \mapsto -A_{m}, G_l\mapsto G_{l}$ both leave \cref{eqn:onsager-algebra} invariant. 

From this structure, it follows that $A_0$ and $A_1$ obey the Dolan-Grady conditions \cite{Dolan82,Perk89,Davies91,Perk16}
\begin{align}
\Big[\big[[A_1,A_0],A_0\big],A_0\Big]=[A_1,A_0] \qquad \qquad
\Big[\big[[A_0,A_1],A_1\big],A_1\Big]=[A_0,A_1] \label{eq:DG}\ .
\end{align}
In fact, \cref{eqn:onsager-algebra,eq:DG} are equivalent \cite{Perk89,Davies91}, and \cref{eq:DG} remains true replacing $A_0\mapsto A_l$, $A_1\mapsto A_m$ \cite{Jones24}.
Ref. \cite{Naudts09} demonstrates that these conditions give rise to operators that we rewrite as

\begin{align} 
    R_{m,l} = \frac{1}{4}A_m + \frac{1}{2}G_{m-l}-\frac{1}{4}A_{2l-m}\ ,
\end{align}

\noindent obeying the commutation relations

\begin{align} \label{eqn:ladder-operators}
    [R_{m,l}, A_l] = R_{m,l} \quad\quad\quad [R_{m,l}^\dag, A_l] = -R_{m,l}^\dag\ .
\end{align}

\noindent We can therefore interpret $R_{m,l}$ as ladder operators for $A_l$; this structure ensures that the spectrum of $A_l$ consists of sectors with unit level spacing.

We can construct $A_0$ and $A_1$ that are periodic, $L$-site, $\mathbb{Z}_N$-symmetric chiral clock chains with local Hilbert space dimension $N$, and that obey the Onsager algebra. We will take the chain length $L$ to be finite. Moreover, for convenience, we will choose $L=0 \mod N$, which we discuss further in Appendix \ref{appendix:kramers-wannier}. 
Let us define the single-site ``shift'' and ``clock'' operators $X_j$ and $Z_j$ obeying $(X_j)^N = (Z_j)^N = 1$ and $X_jZ_k = \omega^{\delta_{jk}} Z_kX_j$, where $\omega = e^{2\pi i/N}$. In the $Z$-diagonal basis, with states $\ket{a}$ defined modulo $N$, we have

\begin{align}
    X_j = \sum_{a_j=0}^{N-1}\ket{a_j-1}\bra{a_j} \quad\quad\quad Z_j=\sum_{a_j=0}^{N-1}\omega^{a_j}\ket{a_j}\bra{a_j}.
\end{align}

\noindent The first two generators of the Onsager algebra, related by Kramers-Wannier duality, are given by 

\begin{align} \label{eqn:A01}
\begin{split}
    A_0 = -\frac{1}{N} \sum_{j=1}^L \sum_{m=1}^{N-1} \alpha_m X_j^m \quad\quad\quad
    A_1 = -\frac{1}{N} \sum_{j=1}^L \sum_{m=1}^{N-1} \alpha_m (Z_{j-1}^\dag Z_j)^m
\end{split}
\end{align}

\noindent for complex couplings $\alpha_m = (1-\omega^m)^{-1}$. In the $Z$-diagonal basis, the eigenstates of $A_1$ are $\{\ket{a_1a_2\dots a_L}\}$ and the eigenstates of $A_0$ are $\{\ket{v_1^{(n_1)}v_2^{(n_2)}\dots v_L^{(n_L)}}\}$ for 

\begin{align}
    \ket{v_j^{(n)}} = \frac{1}{\sqrt{N}}\sum_{a_j=0}^{N-1}\omega^{-na_j}\ket{a_j}\ .
\end{align}

\noindent The ground state of $A_0$ is the unique state $\ket{v_1^{(0)}v_2^{(0)}\dots v_L^{(0)}}$, while $A_1$ has the $N$ ferromagnetic ground states $\{\ket{\tau_a} = \ket{aa\dots a}\}$ for $a \in \{0,1,\dots,N-1\}$. These states all have the same energy $E_0 = -L(N-1)/2N$.

The remaining generators can be found using either the commutation relations \cref{eqn:onsager-algebra} or the pivot procedure described in Ref. \cite{Jones24}, utilising the relation

\begin{align} \label{eqn:pivot}
    e^{\beta A_m} A_l e^{-\beta A_m} = \cosh^2\left(\frac{\beta}{2}\right)A_l + \sinh(\beta) G_{m-l}-\sinh^2\left(\frac{\beta}{2}\right)A_{2m-l} \qquad \beta \in \mathbb{C}\ .
\end{align}
Consequently, all $A_l$ for even (odd) $l$ are related by a unitary transformation to $A_0$ ($A_1$), since

\begin{align} \label{eqn:unitary-pivot}
    A_{2k} = (U_1 U_0)^k A_0(U_0^\dag U_1^\dag)^k, \quad\quad\quad A_{2k+1} = (U_1 U_0)^k A_1(U_0^\dag U_1^\dag)^k, 
    \quad\quad\quad A_{-k} = U_0^{} A_k U_0^\dag
\end{align}
\noindent for $U_m = e^{-i\pi A_m}$. Hence, all of these Hamiltonians have the same ground state energy $E_0$, and the ground state $\ket{\psi_l}$ of $A_l$ is unique ($N$-fold degenerate) for even (odd) $l$. Additionally, by considering the light cone of $(U_1 U_0)^k$, it follows that each $A_m$ has interaction range at most $|m|$; our generators are therefore local operators in this representation. We emphasise that while any $A_l$ or $G_m$ could be calculated using this procedure, for $N>2$, a general closed-form expression is not known; however, for $A_{-1}$ and $A_2$, closed formulae have been determined \cite{Ahn91, Vernier19, Jones24}.

Recall that we work with the family of Hamiltonians of the form 

\begin{align} \label{eqn:general-chiral-clock-family}
    H_A[\{t_m\}, N] = \sum_m t_m A_m \qquad t_m \in \mathbb{R} \ .
\end{align}

\noindent  It is convenient to define the corresponding Laurent polynomial 

\begin{align} \label{eqn:laurent-polynomial-define}
    f(z) = \sum_m t_m z^m\ .
\end{align}

\noindent As shown in Results 1 and 2 of Ref. \cite{Jones21}, for $N=2$, if we can write 

\begin{align} \label{eqn:skeleton-condition}
    f(z)=\pm z^p g(z)^2 \quad\quad\quad \text{with}\quad\quad\quad g(z) = \sum_{k=0}^d s_k z^k
\end{align}

\noindent and some integer $p$, then we can write the ground state as an exact MPS.\footnote{Actually, the $N=2$ model we consider differs by $A_k \mapsto(-1)^{k+1} A_k$ compared to Ref. \cite{Jones21} (where $X$ and $Z$ are also interchanged). The alternating sign can be toggled by the unitary transformation $\prod_j Z_j$ and so we can draw the same conclusion.} Hamiltonians parameterised by such coefficients $\vec{s} = (s_0,s_1, s_2, \dots, s_d)$ are said to ``lie on the skeleton''. In fact, for $N=2$, the ground state can be expressed as an MPS if and only if $f(z)= z^p g(z)^2 h(z)$, where $h(z)$ is any Laurent polynomial obeying $h(z) = h(1/z)$. This is loosely motivated by noting that, as discussed in Section \ref{sec:spectrum}, the spectrum of $H_A[\{t_m\}, N]$ involves terms with coefficients $\sim\sqrt{f(z)f(1/z)}$. For Hamiltonians corresponding to Eq. (\ref{eqn:skeleton-condition}), these terms simplify to $\sim g(z)g(1/z)$; the lack of a square-root is suggestive of simplifications in the analysis. We aim to generalise the results of Ref. \cite{Jones21} by demonstrating that, for $N>2$, Hamiltonians satisfying \cref{eqn:skeleton-condition} will have exact MPS ground states in certain gapped regions. However, we will \emph{not} claim that such a condition is \emph{necessary} for the ground state to be an exact MPS.

\begin{figure}
    \centering
    \includegraphics[]{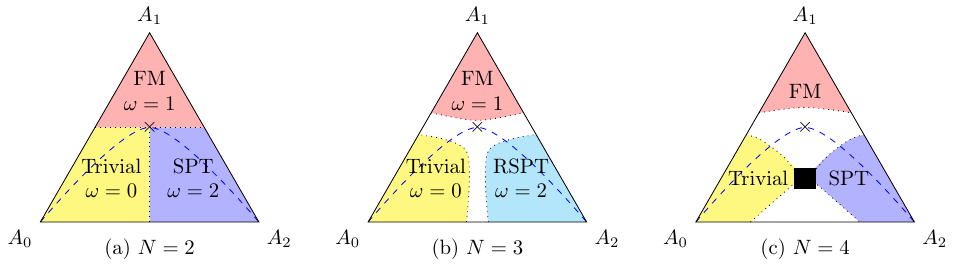}
    \caption{Schematic phase diagrams for the Hamiltonian $H = \alpha A_0+\beta A_1 + \gamma A_2$ with normalisation $\alpha+\beta+\gamma=1$. The winding number of $f(z)$ is denoted by $\omega$. The trivial phase has $\omega=0$, while $\omega=1$ corresponds to the ferromagnetic (FM) phase. For even $N$, $\omega=2$ gives an SPT; for odd $N$, we instead have a representation-SPT (RSPT). The coloured regions are gapped and connected to fixed-point $A_k$, therefore corresponding to the intersection of the set $\mathcal{S}$ with the phase diagram. The white regions are gapless and outside $\mathcal{S}$. For $N=2$, the phase diagram is well-understood \cite{Wolf06, Smith22}. Diagrams for $N=3,4$ were found numerically in Ref. \cite{Jones24}. For $N=4$, it is unclear whether the trivial and SPT phases have a direct transition at the black square. The dashed line represents a leg of the skeleton parameterised by $H = (1-\lambda)^2A_0 + 2\lambda (1-\lambda)A_1 + \lambda^2A_2$. For $N=2$, the ground state is an MPS along the entire path. We show that the analogous state is an eigenstate for general $N$, and is the ground state in the gapped regions. The crosses represent points where the MPS construction given in this paper has a singularity; this coincides with a gap closing in the sector $\mathcal{S}$.}
    \label{fig:phase-diagrams-d=1}
\end{figure}

As illustrated in Fig. \ref{fig:phase-diagrams-d=1}, the $N>2$ phase diagrams have extended gapless regions; this contrasts with the $N=2$ case. We will find that, on the $N>2$ skeletons, we can construct MPS eigenstates analogous to the $N=2$ ground state except at a measure-zero set of points. However, we will find that these eigenstates are only ground states in gapped regions containing fixed-point Hamiltonians $A_l$. We represent these by the coloured regions in Fig. \ref{fig:phase-diagrams-d=1}. It is therefore convenient to define the set $\mathcal{S}$ as the set of Hamiltonians in the class $H_A$ that lie within such gapped regions, as in \cref{eq:S}.

Finally, we note that the structure of the Onsager algebra imposes that 

\begin{align} \label{eqn:An+Gnpsi}
    (A_n+G_n)\ket{\psi_0} \propto \ket{\psi_0} \quad\quad\quad (A_n+G_n)\ket{\phi^{(P)}_{E_0+1/N, \, p=0}} \propto \ket{\phi^{(P)}_{E_0+1/N, \, p=0}}\ ,
\end{align}

\noindent where $\ket{\psi_0}$ is the ground state of $A_0$ and

\begin{align}
    \ket{\phi^{(P)}_{E_0+1/N, \, p=0}} = \frac{1}{\sqrt{L}}\sum_{j=1}^L e^{-i P j} Z_j^\dag \ket{\psi_0} \quad\quad\quad P \in \bigg\{0, \frac{2\pi}{L}, \frac{4\pi}{L} , \dots ,  \frac{2(L-1)\pi}{L}\bigg\}\ .
\end{align}

\noindent We will use these results in Section \ref{subsec:theorem-1-proof}, and they are proved in Appendix \ref{appendix:AplusGpsi} using the ladder operators \cref{eqn:ladder-operators}.
\subsection{Remarks on the phase diagram}\label{sec:phasediagram}
The family of models, $H_A$, has a $D_{2N}=\mathbb{Z}_N\rtimes \mathbb{Z}_2^{\mathrm{CPT}}$ symmetry. The $\mathbb{Z}_N$ generator is $\prod_j X_j$, while the $\mathrm{CPT}$ symmetry combines charge conjugation, lattice inversion, and complex conjugation \cite{Jones24}. For all $N$, the $A_{2k+1}$ are symmetry-breaking phases, with $N$-fold ground-state degeneracy. In the symmetric phases we have a unique ground state, and it is of interest to consider its SPT class \cite{Chen11}. For odd $N$, $D_{2N}$ has no SPT phases, and each $A_{2k}$ is in the trivial symmetric phase. For even $N$, $A_{4k}$ is trivial while $A_{4k+2}$ is a non-trivial $D_{2N}$ SPT; this was shown using charge pumps in Ref.~\cite{Jones25}. For odd $N$, $A_2$ has some SPT-like features, and, while trivial, the nearby region is labelled a representation-SPT (RSPT) \cite{OBrien20,Jones24,Verresen25}, due to a non-trivial linear symmetry-fractionalisation in the leading entanglement levels. 

The above analysis relates to the SPT class when allowing arbitrary symmetric perturbations to our Hamiltonian. In the $N=2$ case, the BDI class has an integer topological invariant, coinciding with the winding number of $f(z)$. This cannot change without a phase transition, provided we stay within the BDI class of free-fermion models \cite{Verresen18}. In fact, the set $\mathcal{S}$ inside the family $H_A$ also has the same integer invariant. As we discuss in \Cref{sec:spectrum}, zeros of $f(z)$ on the unit circle correspond to a bulk gap closing within $\mathcal{S}$. Unlike the $N=2$ case, we can leave $\mathcal{S}$ without a zero crossing the unit circle due to a first order transition to a different sector. However, if we remain in $\mathcal{S}$, we cannot have a zero of $f(z)$ cross the unit circle, and so the winding number is invariant. This is one way to understand the distinct symmetric phases appearing for $N=3$ in \cref{fig:phase-diagrams-d=1}: they have the same SPT invariant, but different winding numbers. Thus, $\mathcal{S}$ consists of disjoint gapped regions around each of the $A_k$, each dual under Kramers-Wannier and unitary pivots.
\subsection{Main results}

Here, we state our main results; these will be proved in Section \ref{sec:analysis}. We characterise exact eigenstates of Hamiltonians on connected networks $f(z) = \pm z^p g(z)^2$ within the family \cref{eqn:general-chiral-clock-family} and show that they are ground states in certain regions of the phase diagram. Moreover, these states can be written in exact MPS form.

To write down the closed form of the wavefunction, we must first compute a collection of coefficients $\{b_k\}$ from the coefficients $\{s_k\}$ of $g(z) = \sum_k s_k z^k$. This can be done algorithmically \cite{Jones21}, which will be discussed in Section \ref{subsec:theorem-1-proof}. We note that this algorithm is identical to the Schur-Cohn algorithm \cite{Schur18, Cohn22}, which is used to determine the distribution of the roots of a polynomial with respect to the unit circle. Furthermore, we require the fixed-point ground states $\ket{\psi_p^\pm}$ of $\pm A_p$. For the clock models, these can be written in exact MPS form using the procedure given in Ref. \cite{Jones24}, and in the symmetry-breaking case the result holds for any of the $N$ symmetric ground states $\ket{\psi_p^{(i)}}$.

\begin{shaded}
 \begin{result}[Exact eigenstate]\label{result:theorem-1}
If we can express the Laurent polynomial in the form $f(z) = \pm z^p g(z)^2$ for $p \in \mathbb{Z}$ and 

\begin{align}
    g(z) = \sum_{k=0}^d s_k z^k\ ,
\end{align}

\noindent then the corresponding Hamiltonian has the eigenstate

\begin{align} \label{eqn:gs-conjecture-varphi}
    \ket{\varphi} = \mathrm{M}^{(d)} \mathrm{M}^{(d-1)}\dots \mathrm{M}^{(1)} \ket{\psi_p^\pm}\ .
\end{align}

\noindent Here, $\mathrm{M}^{(k)}=\exp(\mp \beta_k A_{p+k})$ for $\beta_k = 2 \textup{ arctanh}(b_k)$, and $\ket{\psi_p^\pm}$ is the ground state of the Hamiltonian $\pm A_p$. Note this breaks down if any $|b_k|=1$.

\end{result}\end{shaded}
\noindent The states $\ket{\varphi}$ can be expressed in MPS form,\footnote{The state $\ket{\varphi}$ is a series of finite-bond dimension MPOs $\mathrm{M}^{(k)}$ operating on an MPS $\ket{\psi_p}$, and so it must itself be an MPS of finite bond dimension. Upper bounds on the bond dimension are straightforward to find, since the $A_m$ have support on $|m|+1$ sites. For example, for $p=0$, an upper bound on the bond dimension $\chi$ is $\log_N\chi=\frac{1}{2}d(d+3)$. Importantly, this does not depend on system size. For the $N=2$ case, optimal bounds on the bond dimensions can be found in certain cases, see discussion in Refs.~\cite{Jones21,Jones23}.} and these Hamiltonians therefore constitute the MPS skeleton; however, a measure-zero set of points, where any $|b_k|=1$, must be excluded. We prove this result in Section \ref{subsec:theorem-1-proof}. 

\begin{shaded}
 \begin{result}[Exact ground state]\label{result:theorem-2} For all vectors $\vec{s} = (s_0, \dots, s_d)$ that do not give any $|b_k|=1$, the eigenstate $\ket{\varphi}$ is the ground state of $H_A(\vec{s})$ when $H_A(\vec{s})\in \mathcal{S}$.
\end{result}\end{shaded}

\noindent This follows straightforwardly from the perturbative arguments we provide in Section \ref{subsec:theorem-2-proof}. Note that $|b_k|=1$ occurs in gapless models (see Appendix \ref{appendix:schur-cohn-algorithm}), but can also arise for Hamiltonians in $\mathcal{S}$. In such cases, the method used to construct the MPS breaks down, but we can argue that we have an exact MPS ground state by considering a limit of models, lying in $\mathcal{S}$ and on the skeleton, where we can explicitly construct the ground state; Refs. \cite{Jones21,Jones23} give related discussions. In the $N=2$ case, the distinct methods of Ref. \cite{Jones23} prove the existence of MPS ground states for $f(z)=\pm z^p g(z)^2$ with no restrictions on the $\{b_k\}$.

Although we cannot generally express $f(z)=\pm z^p g(z)^2$ for a finite polynomial $g(z)$, we can always define and truncate an appropriate ``square-root'' series  at some finite power $D$ of $z$ \cite{Jones21,Jones23}. This gives an approximate Hamiltonian $H_A^{(D)}$ corresponding to the Laurent polynomial $f_D(z) = \pm z^k g_D(z)^2$ (note $k\neq p$ in general). For example, the chiral Potts model $A_0 + \lambda A_1$, has Laurent polynomial $f(z) = 1+\lambda z$. This can be approximated by \begin{align}f_{3}(z) = g_{3}(z)^2 = 1+\lambda z + \frac{5}{64} \lambda^4 z^4 - \frac{1}{64}\lambda^5 z^5+\frac{1}{256}\lambda^6 z^6 \ ,\end{align} for $g_{3}(z) = 1+\lambda z/2 -\lambda^2z^2/8+\lambda^3z^3/16$. Note that $H_A^{(3)}$ is then the chiral Potts model, plus a ``small'' perturbation by longer-range models. The ground state of the Hamiltonian $H_A^{(D)}$ can be used to approximate the ground state of $H_A$. Indeed, the skeleton constitutes a dense subset of our family of Hamiltonians in the following sense.

\begin{shaded}
 \begin{result}[The skeleton is dense]\label{result:theorem-3} Consider a general chiral clock Hamiltonian $H_A$ of the form \cref{eqn:general-chiral-clock-family}, with ground state $\ket{\varphi}$, and let 
 $\mathcal{E}_A$ be the corresponding ground state energy density of  $H_A^{}$. We can define a Hamiltonian $H_A^{(D)}$ on the skeleton with ground state $\ket{\varphi_D}$, such that for any fixed $\eps>0$ and sufficiently large $D$ and $L$, we find that the energy density $\mathcal{E}_A^{(D)}$ of $\ket{\varphi_D}$ with respect to $H_A^{}$ satisfies $\lvert \mathcal{E}_A^{(D)}-\mathcal{E}_A^{}\rvert < \eps$. Analogous statements hold for the ground state in any fixed sector.
\end{result}\end{shaded}
\noindent 
The proof is given in Section \ref{subsec:theorem-3-proof}.
Heuristically, this tells us that the ground state of any Hamiltonian in $\mathcal{S}$ can be approximated by a sequence of states of the form \cref{eqn:gs-conjecture-varphi}. 
Ref.~\cite{Jones21} gives a derivation of the scaling dimension of the disorder operator in the Ising conformal field theory using such a path of skeleton states.
We expect that this can be made precise in the thermodynamic limit using the results of \cite{Bratelli78}.

\begin{shaded}
 \begin{result}[More exact eigenstates]\label{result:theorem-1.1}
For even $p$, Hamiltonians with Laurent polynomials $f(z) = z^p g(z)^2$ have the exact eigenstates

\begin{align} \label{eqn:excited-state-conjecture}
    \ket{\chi_p^{(P)}} = \mathrm{M}^{(d)} \mathrm{M}^{(d-1)}\dots \mathrm{M}^{(1)} \ket{\phi^{(P)}_{E_0+1/N,\,p}}\ ,
\end{align}

\noindent where 

\begin{align}
    \ket{\phi^{(P)}_{E_0+1/N, \,p}} = \frac{1}{\sqrt{L}}\sum_{j=1}^L e^{-i P j} (U_1 U_0)^{p/2} Z_j^\dag \ket{\psi_0^+}, \quad\quad\quad P \in \bigg\{0, \frac{2\pi}{L}, \frac{4\pi}{L} , \dots ,  \frac{2(L-1)\pi}{L}\bigg\}\ .
\end{align}

\noindent For $f(z) = -z^p g(z)^2$ and even $p$, an equivalent set of eigenstates exist where we replace $Z_j^\dag \mapsto Z_j$ and $\ket{\psi_0^+} \mapsto \ket{\psi_0^-}$.

\end{result}\end{shaded}

\noindent This result is proved in Section \ref{subsec:theorem-4-proof}, where we also discuss the necessity of even $p$ in $f(z)=\pm z^pg(z)^2$ for this to hold. These states can also be represented in exact MPS form.

\subsection{Further remarks}\label{remark}
We note that \cref{result:theorem-1} can be proved if our model obeys the Onsager algebra and when $\ket{\psi_p^\pm}$ is unique.\footnote{While this is not the case in the clock models for odd $p$, we actually only require that the ground state eigenspace of $A_p$ has a basis that diagonalises $A_\alpha+G_\alpha$ for all $\alpha$. While guaranteed if the ground state is unique (as for even $p$), a suitable basis can also be constructed for odd $p$ in the clock models, as discussed in Appendix \ref{appendix:kramers-wannier}.} For $\ket{\varphi}$ to have an exact MPS representation, we additionally require that $A_0$ and $A_1$ have exact MPS ground states and are finite-range. Moreover, \cref{result:theorem-2} requires that our model obeys a finite-dimensional Onsager algebra and that each $A_p$ is gapped. \cref{result:theorem-3} holds generally for any finite-dimensional representation of the Onsager algebra, provided that the energy density remains finite in the thermodynamic limit. \cref{result:theorem-1.1}, however, is specific to the clock models. Finally, for brevity, we will henceforth write $\ket{\psi_p^+}$ as $\ket{\psi_p}$.

We remark that, just as the $N=2$ case can be viewed as a family of free-fermion models, for $N>2$, our family $H_A$ has a dual parafermionic form \cite{Fradkin80,Fendley12}. Many of our results will carry over simply to the parafermion picture, although care must be taken with defining parafermionic MPS \cite{Xu17}. Following our discussion about the winding number in \cref{sec:results}, it would be interesting to understand if there is a connection between this winding and the number of edge modes \cite{Fendley12}. 

\section{\label{sec:examples}Examples}
In this section, we motivate our results by deriving \cref{result:theorem-1}, \cref{result:theorem-2}, and \cref{result:theorem-1.1} explicitly for the simplest non-trivial case. We then demonstrate how the ground state can be expressed as an exact MPS. Finally, we show how this result can be used to evaluate the disorder operator on the skeleton, generalising known results for $N=2$ to any even integer $N$.

\subsection{Eigenstates for \texorpdfstring{$d=1$}{d=1}}
Here, we illustrate our results in the case of $d=1$ using a variation of the method used in the proof for general $d$. The initial steps of the argument are illustrative of the general method, but we are able to make certain computations explicit in this particular case. We also utilise a known expression for the ground state energy in $\mathcal{S}$ that bypasses the perturbative argument of \Cref{result:theorem-2}. 

Let us consider the $d=1$ Hamiltonian

\begin{align}
    H_A = A_{0} + 2 a A_1 + a^2 A_2\ ,
\end{align}

\noindent which has $p=0$ and $g(z) = 1+az$. The non-interacting $N=2$ case of $H_A$ was studied in detail in Refs. \cite{Smith22, Wolf06}. 

Defining $\tilde{H}_A =  e^{\beta A_1}H_A e^{-\beta A_1}$ for $\beta = 2\text{ arctanh}\left(a\right)$, the pivot relation \cref{eqn:pivot} gives

\begin{align}
    \tilde{H}_A = (a^2+1)A_{0} + 2a(A_1+G_1)\ .
\end{align}

\noindent Consider the ground state $\ket{\psi_0}$ of $A_0$, which---in the chiral clock representation---can be written as the product state \cite{Jones24}

\begin{align}
    \ket{\psi_0} = \ket{v_1^{(0)}v_2^{(0)}\dots v_L^{(0)}} \quad\quad\quad\text{where}\quad\quad\quad \ket{v_j^{(0)}} = \frac{1}{\sqrt{N}} \sum_{a_j=0}^{N-1} \ket{a_j}\ ,
\end{align}
and the translationally invariant basis of single-particle excitations $\{\ket{\phi_{E_0+1/N,\, p=0}^{(P)}}\}$ labelled by momentum $P$. We have 

\begin{align}
    A_{0} \ket{\psi_{0}} = E_0\ket{\psi_{0}} \quad\quad\quad A_{0} \ket{\phi_{E_0+1/N,\, p=0}^{(P)}} = \left(E_0+\frac{1}{N}\right)\ket{\phi_{E_0+1/N,\, p=0}^{(P)}}\ .
\end{align}

\noindent Moreover, from \cref{eqn:An+Gnpsi}, we have 

\begin{align}
    (A_1+G_1) \ket{\psi_{0}} \propto \ket{\psi_0} \quad\quad\quad(A_1+G_1) \ket{\phi_{E_0+1/N,\, p=0}^{(P)}} \propto \ket{\phi_{E_0+1/N,\, p=0}^{(P)}}\ .
\end{align}

\noindent The states $\ket{\psi_{0}}$ and $\ket{\phi_{E_0+1/N,\, p=0}^{(P)}}$ are therefore eigenstates of $\tilde{H}_A$. In fact, we can show $(A_1+G_1) \ket{\psi_{0}} = 0$ by noting that there exists an operator $\mathcal{W}$ satisfying both $\mathcal{W}\ket{\psi_0}=\ket{\psi_0}$ and $\{\mathcal{W}, A_{2k+1}\}=0$ \cite{Albertini89b}.\footnote{$\mathcal{W}$ is only defined for $L = 0 \mod N$. However, for any chain length, we can show $(A_1+G_1) \ket{\psi_{0}} = 0$ by explicitly evaluating over the lattice.} Thus, reversing the transformation, we have

\begin{align}
    H_A\ket{\varphi} = (a^2+1) E_0 \ket{\varphi} \quad\quad\quad \text{and} \quad\quad\quad H_A\ket{\chi_{p=0}^{(P)}} \propto \ket{\chi_{p=0}^{(P)}}
\end{align}

\noindent for 

\begin{align}
    \ket{\varphi} = e^{-\beta A_1}\ket{\psi_0}, \quad\quad\quad \ket{\chi_{p=0}^{(P)}} = e^{-\beta A_1} \ket{\phi_{E_0+1/N,\, p=0}^{(P)}}\ .
\end{align}

Therefore, for all $a$, $\ket{\varphi}$ and $\{\ket{\chi_{p=0}^{(P)}}|P\in \{0, 2\pi/L,\dots,2(L-1)\pi/L\}\}$ are eigenstates of $H_A$. The state $\ket{\varphi}$ has energy $\varepsilon_\varphi=(a^2+1) E_0$. From Refs. \cite{Davies90, Ahn91}, we know that this is the ground state energy provided $H_A \in \mathcal{S}$. Thus, for $d=1$, \cref{result:theorem-1} and \cref{result:theorem-2} hold. 

\subsection{Ground state as an MPS for \texorpdfstring{$d=1$}{d=1}}

\begin{figure}
    \centering
    \mbox{
    \includegraphics[]{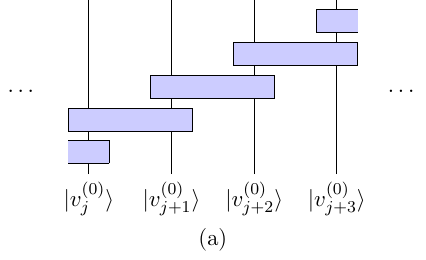}
    \hspace{1cm}
    \includegraphics[]{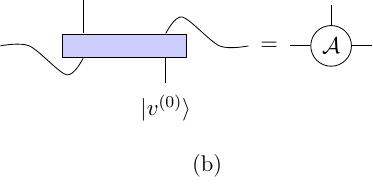}
    }
    \label{fig:MPS-tensor}
    \caption{(a) The circuit diagram for the $d=1$ ground state. This can be interpreted as an MPO acting on a product state, giving an MPS. The blue boxes represent each unitary gate $U_{j,j+1}$, and the $\ket{v_j}$ are the single-site states. (b) The MPS tensor from \cref{eqn:MPS-tensor-d=1}.}
    \label{fig:MPS-circuit}
\end{figure}

Let us write 

\begin{align}
    e^{-\beta A_1} 
    = \prod_{j=1}^L U_{j, j+1}(\beta)\ ,
\end{align}

\noindent where

\begin{align}
    U_{j, j+1}=\exp \left( \frac{\beta}{N}  \sum_{m=1}^{N-1} \alpha_m (Z_{j}^\dag Z_{j+1}^{})^m\right)
\end{align}

\noindent are commuting operators. The ground state $\ket{\varphi}$ can therefore be represented by the circuit in Fig. \ref{fig:MPS-circuit}. This can be interpreted as a matrix-product operator (MPO) acting on a product state, giving an MPS. We obtain the (unnormalised) MPS form

\begin{align}
\begin{split}
    \ket{\varphi} &=  \sum_{a_1 \dots a_L = 0}^{N-1}\left( \prod_{j=1}^L U_{j-1, j}(\beta) \right) \ket{a_1 a_2 \dots a_L}\\
    &= \sum_{a_1 \dots a_L = 0}^{N-1} \text{Tr}\left[\mathcal{A}^{(a_1)}_1 \mathcal{A}^{(a_2)}_2 \dots \mathcal{A}^{(a_L)}_L\right]\ket{a_1 a_2 \dots a_L}
\end{split}
\end{align}

\noindent with the MPS tensors

\begin{align}\label{eqn:MPS-tensor-d=1}
\begin{split}
    \left(\mathcal{A}_j^{(a_j)}\right)_{\sigma_{j-1}, \sigma_j} &= \delta_{\sigma_{j-1} a_j} \exp \left(\frac{\beta}{N}\sum_{m=1}^{N-1} \alpha_m \omega^{-m(a_{j}-\sigma_{j})}\right)\\
    &=\delta_{\sigma_{j-1} a_j} \exp \left(\frac{\beta(N-1)-2\beta((a_j-\sigma_j)\mod N)}{2N}\right)
\end{split}
\end{align}

\noindent of bond dimension $\chi = N$.

\subsection{Disorder operator on the \texorpdfstring{$d=1$}{d=1} skeleton} \label{sec:string-order} 

For even $N$, we define the unitary disorder operator

\begin{align} \label{eqn:string-order-operator}
    \mu_k  = \prod_{j=-\infty}^k X_j^{N/2}\ .
\end{align}

\noindent For $N=2$, and for $g(z)$ such that all $|b_k|<1$, the expectation of the disorder operator in the ground state is \cite{Jones21}

\begin{align}
    \braket{\mu_k}_{N=2} = \prod_{l=1}^d (1-b_l^2)^{l/2}\ .
\end{align}

\noindent For $g(z)=1+az$, with $|a|<1$, this becomes $\braket{\mu_k}_{N=2} = \sqrt{1-a^2}$. Here, we outline how this $d=1$ result generalises to any even $N$. Full details of the calculation are given in Appendix \ref{appendix:string-order-details}.

We aim to evaluate\footnote{Note that we evaluate the one-point function for simplicity, ignoring issues with the left boundary. A calculation of the two-point correlation function $\braket{\mu_0\mu_k}$ would avoid such issues, and can be evaluated similarly, leading to the square of \cref{eq:1pt}.}

\begin{align}
    \braket{\mu_k} = \frac{\bra{\varphi}\mu_k\ket{\varphi}}{\braket{\varphi|\varphi}} = \frac{\bra{\psi_0}e^{-\beta A_1}\mu_ke^{-\beta A_1}\ket{\psi_0}}{\braket{\psi_0|e^{-2\beta A_1}|\psi_0}}\ .
\end{align}

\noindent If we define

\begin{align}
    U_{j,j+1}^{(m)}(\beta) = \exp\left(\frac{\beta}{N}\alpha_m(Z_{j}^\dag Z_{j+1}^{})^m\right)\ ,
\end{align}

\noindent then we can use the commutation relations $X_j Z_k = \omega^{\delta_{jk}}Z_k X_j$ and the fact that $\mu_k \ket{\psi_0} = \ket{\psi_0}$ to express the numerator as

\begin{align}
    \bra{\psi_0}e^{-\beta A_1}\mu_ke^{-\beta A_1}\ket{\psi_0}=\bra{\psi_0}e^{-2\beta A_1} \prod_{\substack{m=1 \\ m \text{ odd}}}^{N-1}U_{k,k+1}^{(m)}(-2\beta) \ket{\psi_0}\ .
\end{align}

\noindent If we write $\ket{\psi_0}$ in terms of the states $\ket{a}$ that are diagonal in the $Z$-basis

\begin{align}
    \ket{\psi_0} = \frac{1}{\sqrt{N^L}}\sum_{a_1\dots a_L =0}^{N-1} \ket{a_1 \dots a_L}\ ,
\end{align}

\noindent then $\braket{\mu_k}$ can be written as the ratio of two sums over configurations of states in the $Z$-basis. We can evaluate these expressions with a transfer matrix method, writing 

\begin{align}
    \braket{\mu_k} = \frac{\text{Tr}(T^{L-1} B)}{\text{Tr}(T^L)}
\end{align}

\noindent for two matrices $T$ (the transfer matrix) and $B$. We find that these matrices are circulant, and so their spectra can be determined from standard results \cite{Davis79}. In the thermodynamic limit $L\rightarrow \infty$, the largest eigenvalue of $T$ dominates, and we obtain

\begin{align}
    \braket{\mu_k} = \sqrt{1-a^2} \label{eq:1pt}
\end{align}

\noindent for all even $N$, as desired.

\section{\label{sec:spectrum}The spectrum of Onsager-integrable chiral clock chains}
In this section, we summarise how the Onsager algebra can be used to examine the structure of the spectrum. As stated in Result \ref{result:theorem-2}, the eigenstate $\ket{\varphi}$ is the ground state of Hamiltonians in the region $\mathcal{S}$, defined by Eq. (\ref{eq:S}). As this region is determined by the spectral gap of the Hamiltonians, it is important to understand some properties of their spectra. The reader may wish to skip this section on a first read, noting only Eq. (\ref{eqn:spectrum-form}) which is used in Section \ref{subsec:constraints-on-ralpha}.

Firstly, we note that chiral clock Hamiltonians commute with the translation operator and the $D_{2N}=\mathbb{Z}_N\rtimes\mathbb{Z}_2^{\mathrm{CPT}}$ generators described in \Cref{sec:results}.
\noindent The Hamiltonian can therefore be block-diagonalised into sectors labelled by the quantum numbers of momentum and $\mathbb{Z}_N$-charge, denoted $P$ and $Q$ respectively. 
Additionally, as discussed in Refs. \cite{Davies90, Roan91,McCoy91,VonGehlen01,Roan11}, the Hamiltonians can be further block-diagonalised into Onsager sectors, with multiple Onsager sectors per $(P,Q)$ sector. We define the ``ground-state Onsager sector'', lying within the $(0,0)$ sector, to be the Onsager sector containing the ground state of $A_0$.

As discussed in Refs. \cite{Davies90, Ahn91}, given a finite-dimensional representation of the Onsager algebra, it is possible to write $\sum_{m=-n}^n \alpha_m A_m = 0$ for some integer $n$ and coefficients satisfying $\alpha_m = \pm \alpha_{-m}$; determining these coefficients is, however, non-trivial. Nevertheless, one can define $z_{\mp k} = e^{\pm i m \theta_k }$ as the roots of the polynomial $\mathfrak{h}(z) = \sum_{m=-n}^n \alpha_m z^{m+n}$. The generators of any finite-dimensional representation of the Onsager algebra can then be rewritten, up to sector-dependent additive constants, with the invertible transformation\footnote{Note that a slight modification must be made if $\mathfrak{h}(z)$ has repeated roots. As noted in Ref. \cite{Davies90}, it is not necessary to consider this case for the clock models.}

\begin{align} \label{eqn:su2-transformation}
    A_m = \frac{1}{2} \sum_{k=1}^n e^{-im\theta_k} E_k^+ + e^{im\theta_k}E_k^-\quad\quad\quad
    G_m = \frac{1}{2}\sum_{k=1}^n \left( e^{-im\theta_k} -  e^{im\theta_k}\right) H_k\ ,
\end{align}

\noindent where the Onsager algebra imposes that

\begin{align}
    [E_j^+,E_k^-]=\delta_{jk}H_k\quad\quad\quad
    [H_j,E_k^\pm]=\pm2\delta_{jk}E_k^\pm\ .
\end{align}

\noindent The operators $\{E_j^\pm, H_k\}$ can therefore be interpreted as $sl(2)$ generators. The roots of $\mathfrak{h}(z)$ lie on the unit circle to ensure the self-adjointness of our Hamiltonian, and they are also sector-dependent.

In general, determining the $n$ values $\{\theta_k\}$ is difficult and must be done by solving functional equations \cite{Baxter88,Baxter89,Albertini89b,Davies90, Dasmahapatra93}. (These are referred to as BAMP polynomials in Ref. \cite{Iorgov10} and chiral Potts polynomials in Ref. \cite{Nishino06}.) However, without calculating the values $\theta_k$, we can easily establish the \emph{structure} of the spectra of Onsager-integrable Hamiltonians by writing

\begin{align}
    H_A = \sum_m t_m A_m = \sum_m \sum_{k=1}^n t_m \bigg[\cos (m\theta_k) J_x^k + \sin (m\theta_k) J_y^k\bigg]\ ,
\end{align}

\noindent where we have defined $E^\pm_k = J_x^k \pm i J_y^k$. This can be diagonalised by rotating in the $(J_x,J_y)$ plane to give 

\begin{align} 
\begin{split}
    H_A &= \sum_{k=1}^n \sqrt{\sum_{l,m}t_l t_m \cos\left((m-l)\theta_k\right)} J^k_{x, \text{new}} \\&= \sum_{k=1}^n \sqrt{f(e^{i\theta_k})f(e^{-i\theta_k})}J^k_{x, \text{new}}\\
    &= \sum_{k=1}^n g(e^{i\theta_k}) g(e^{-i\theta_k}) J^k_{x, \text{new}}\ ,
\end{split}\label{eq:spectrum}
\end{align}

\noindent where the final line follows only if we are on the skeleton, where $f(z)=\pm z^p g(z)^2$. Given that we know the eigenvalues of the spin-operators $J^k_{x, \text{new}}$, the form of the spectrum in each sector, up to additive constants, is

\begin{align}\label{eqn:spectrum-form}
\begin{split}
    \varepsilon_A(\{m_k\}) &= \sum_{l,m}s_l s_m\sum_{k=1}^n m_k  \cos((l-m)\theta_k)
\end{split} \ .
\end{align}

\noindent Here, each of the $m_k$ can take values from $\{-S, \dots, S-1, S\}$ for the spin-$S$ representation of $sl(2)$; in the clock models we have $S=1/2$ \cite{Davies90}. Again, we emphasise that this form is up to sector-dependent additive constants. 

For chiral-clock Hamiltonians $H_A\in \mathcal{S}$, the ground state lies in the ``ground-state Onsager sector''. This follows from perturbative arguments and pivoting and duality relations. In this sector, the $\{\theta_k\}$ values can be calculated \cite{Baxter88, Baxter89, McCoy91} and have multiplicity one.\footnote{Generally speaking, computing eigenvalues from functional equations does not give information about their multiplicity, which can be zero \cite{Baxter89}.} The ground state energy density for chiral clock Hamiltonians $H_A \in \mathcal{S}$ in the thermodynamic limit is then given by \cite{Baxter88}

\begin{align}
    \varepsilon_0 = -\frac{1}{2\pi}\sum_{l,m} s_l s_m \int_0^{\pi(1-1/N)} \cos\Big((l-m)\theta\Big)\,\mathrm{d}\zeta
\end{align}

\noindent where

\begin{align}
    \tan(\theta/2) = \left(\frac{\sin (\zeta)}{\sin (\zeta+\pi/N)}\right)^{N/2}
\end{align}

\noindent and $0\leq\theta\leq \pi$. Since we have $f(e^{\pm i \theta})$ in \cref{eq:spectrum}, this range of $\theta$ covers the unit circle. Thus \cref{eq:spectrum} connects gapless modes in this sector to zeros of $f(e^{i \theta})$ on the unit circle. This means that, even when we do not have a first-order transition to a different sector, if we tune a zero to the unit circle, $H_A$ cannot remain in $\mathcal{S}$. This is consistent with previous analysis of the superintegrable chiral Potts line \cite{Cardy93}. 

As mentioned above, some results for the eigenvectors are also known \cite{Albertini89b,McCoy91,AuYang08,AuYang09,Roan11}. These are typically given for the model $A_0 + \lambda A_1$, but we expect they could be generalised. However, this would correspond in the $N=2$ case to knowledge of the diagonal free-fermionic modes. For $N=2$, going from the mode representation to the MPS representation is non-trivial \cite{Jones21}, and our simple construction of the MPS eigenstates in these interacting models gives a new perspective and a direct approach that is mostly decoupled from the exact solutions above.

\section{\label{sec:analysis}Analysis}
In this section, we prove the results given in Section \ref{sec:results}. Given that Hamiltonians with $f(z) = \pm z^p g(z)^2$ are all related to the Hamiltonian with $f(z) = \pm g(z)^2$ via some combination of Kramers-Wannier duality and unitary transformations, we first prove our results for $p=0$.

\subsection{Proof of Result 1} \label{subsec:theorem-1-proof}

We begin by noting that the pivot relation \cref{eqn:pivot}, following directly from the Onsager algebra, means that abstractly

\begin{align} \label{eqn:generalHtilde=aA+gG}
\begin{split}
    \tilde{H}_A &= e^{\beta_1 A_1} \dots e^{\beta_d A_d} H_A e^{-\beta_d A_d} \dots e^{-\beta_1 A_1}\\
    &= \sum_\alpha a_\alpha A_\alpha + g_\alpha G_\alpha
\end{split}
\end{align}

\noindent has coefficients $\{a_\alpha, g_\alpha\}$ independent of our choice of representation. Therefore, if we find the explicit form of $\tilde{H}_A$ in one representation, the form must hold for all representations provided we do not have any representation-specific cancellations of the generators\footnote{For example, for $N=2$, we have $A_l = \pm A_{l+L}$ and $G_l = \pm G_{l+L}$. These terms could cause cancellations which would not occur in models with $N>2$, leading us to conclude the wrong general form of $\{a_\alpha, g_\alpha\}$. We will argue that such cancellations do not occur in our calculations.}.

Thus, we will study a particular (free-fermion) representation below. Define the Majorana fermions

\begin{align}
    \gamma_j = \left(\prod_{k=1}^{j-1}X_k\right) Y_j \quad\quad\quad \tilde{\gamma}_j = \left(\prod_{k=1}^{j-1}X_k\right) Z_j\ ,
\end{align}

\noindent where $X_j, Y_j, Z_j$ reduce to the usual $2\times 2$ Pauli matrices. A representation of the Onsager algebra has generators

\begin{align}
    A_k = \frac{1}{4}\sum_{n=1}^L i \tilde{\gamma}_n \gamma_{n+k} \quad\quad\quad G_k = \frac{1}{8}\sum_{n=1}^L  \tilde{\gamma}_n\tilde{\gamma}_{n+k} -\gamma_{n}\gamma_{n+k}\ .
\end{align}
\noindent This is a slightly different $N=2$ representation of the Onsager algebra, related to our definition of the $N=2$ chiral clock model by a unitary transformation (see \cref{subsec:chiral-clock-intro}). Using results from Ref. \cite{Jones21}, Hamiltonians in this representation with $f(z) = z^p g(z)^2$ can be expressed as 

\begin{align}
    H_A = \frac{1}{8} \sum_{n=1}^L \left(\Omega_n\Gamma_n - 2|\vec{s}|^2\right)\ ,
\end{align}

\noindent where 

\begin{align}
    \Gamma_n = \sum_{\alpha=0}^d s_\alpha\left(\gamma_{n+\alpha} - i \tilde{\gamma}_{n-\alpha}\right) \quad\quad\quad \Omega_n = \sum_{\alpha=0}^d s_\alpha\left(\gamma_{n+\alpha} + i \tilde{\gamma}_{n-\alpha}\right)\ .
\end{align}

\noindent In contrast with the approach taken in Ref. \cite{Jones21}, we treat $\Omega_n$ and $\Gamma_n$ as separate objects rather than writing $\Omega_n = \Gamma_n^\dag$ and considering only transformations on $\Gamma_n$.

Begin by defining $\mathrm{M}^{(k)}$ as in \cref{result:theorem-1}. We wish to compute

\begin{align} \label{eqn:Htilde-definition-rotations}
    \tilde{H}_A = \mathrm{M}^{(1)^{-1}}\dots\mathrm{M}^{(d)^{-1}} H_A \mathrm{M}^{(d)}\dots \mathrm{M}^{(1)}\ ,
\end{align}

\noindent which can be done by taking

\begin{align}
    \Gamma_n \mapsto \tilde{\Gamma}_n = \mathrm{M}^{(1)^{-1}}\dots\mathrm{M}^{(d)^{-1}} \Gamma_n \mathrm{M}^{(d)}\dots \mathrm{M}^{(1)} 
\end{align}

\begin{align}
    \Omega_n \mapsto \tilde{\Omega}_n = \mathrm{M}^{(1)^{-1}}\dots\mathrm{M}^{(d)^{-1}} \Omega_n \mathrm{M}^{(d)}\dots \mathrm{M}^{(1)}\ .
\end{align}

\noindent To evaluate these objects, we observe that 

\begin{align}
    \mathrm{M}^{(k)^{-1}} \gamma_n \mathrm{M}^{(k)} =\frac{\gamma_n+ib_k\tilde{\gamma}_{n-k}}{\sqrt{1-b_k^2}} \quad\quad\quad \mathrm{M}^{(k)^{-1}} \tilde{\gamma}_n \mathrm{M}^{(k)} = \frac{\tilde{\gamma}_n-ib_k\gamma_{n+k}}{\sqrt{1-b_k^2}}\ ,
\end{align}

\noindent which, provided all $|b_k| \neq 1$, gives 

\begin{align} \label{eqn:layer-operation}
    \mathrm{M}^{(k)^{-1}}\left(\sum_{\alpha=-m}^d s_\alpha\left(\gamma_{n+\alpha}\pm i \tilde{\gamma}_{n-\alpha}\right)\right) \mathrm{M}^{(k)} \nonumber  \propto \sum_{\alpha=-m}^d s_\alpha\left(\gamma_{n+\alpha}\pm i \tilde{\gamma}_{n-\alpha}\right) \nonumber  \pm b_k \sum_{\alpha = k-d}^{k+m} s_{k-\alpha} \left(\gamma_{n+\alpha}\pm i \tilde{\gamma}_{n-\alpha}\right)\ .
\end{align}

Define Algorithm \ref{alg:compute-bk} \cite{Jones21} for the coefficients $\{b_k\}$. This is designed such that each layer of the algorithm acting on $\Gamma_n$ eliminates the term $\left(\gamma_{n+\alpha}-i \tilde{\gamma}_{n-\alpha}\right)$ with the largest $\alpha$. As mentioned, this coincides with the Schur-Cohn algorithm \cite{Schur18, Cohn22}.

\begin{algorithm}[H]
\caption{Compute $\{b_k\}$ from vector $\vec{s} = (s_0, \dots, s_d)$}\label{alg:compute-bk}
\begin{algorithmic}
\Require Vector $\vec{s} = (s_0, \dots, s_d)$ with $s_0 \neq 0$
\Ensure Returns the vector $(b_1, \dots, b_d)$

\State Initialize empty list \texttt{bList} $ \gets ()$
\For{$k = d, \dots,2, 1$}
    \State $b_k \gets s_k/s_0$
    \State Append $b_k$ to \texttt{bList} 
    \For{$i = 0, \dots, k$}
        \State $s_i \gets s_i - b_k \cdot s_{k - i}$
    \EndFor
\EndFor
\State \Return Vector \texttt{bList}
\end{algorithmic}
\end{algorithm}

\noindent To demonstrate this, note that

\begin{align}
\begin{split}
    \mathrm{M}^{(d)^{-1}}{\Gamma}_n \mathrm{M}^{(d)} &\propto \sum_{\alpha=0}^d (s_\alpha - b_d s_{d-\alpha})\left(\gamma_{n+\alpha}-i \tilde{\gamma}_{n-\alpha}\right)\\
    &\propto \sum_{\alpha=0}^{d-1} s_\alpha^{(d)}\left(\gamma_{n+\alpha}-i \tilde{\gamma}_{n-\alpha}\right)\ ,
\end{split}
\end{align}

\noindent where $s_\alpha^{(d)} = s_\alpha - b_d s_{d-\alpha}$. This leads to the recursion

\begin{align}
    \mathrm{M}^{(k)^{-1}} \dots \mathrm{M}^{(d)^{-1}}{\Gamma}_n  \mathrm{M}^{(d)}\dots \mathrm{M}^{(k)}  \propto \sum_{\alpha=0}^{k-1} s_\alpha^{(k)}\left(\gamma_{n+\alpha}-i \tilde{\gamma}_{n-\alpha}\right)\ , 
\end{align}

\noindent where $s_\alpha^{(k)} = s_\alpha^{(k+1)} - b^{}_k s_{k-\alpha}^{(k+1)}$ for initial condition $s_\alpha^{(d+1)} = s_\alpha$. After all $d$ layers have been applied, this collapses onto $\tilde{\Gamma}_n \propto \gamma_{n}-i \tilde{\gamma}_{n}$. No terms, however, are necessarily eliminated in $\Omega_n$, and each layer instead generates a new term $\gamma_{n+\alpha}+i \tilde{\gamma}_{n-\alpha}$ with $\alpha = k-d$.

Therefore, for $N=2$, the transformation leads to

\begin{align} \label{eqn:Htilde-Majoranas}
\begin{split}
    \tilde{H}_A &= \frac{1}{2}\sum_{n=1}^L\left(\sum_{\alpha=1-d}^d r'_\alpha \left(\gamma_{n}-i \tilde{\gamma}_{n}\right)\left(\gamma_{n+\alpha}+i \tilde{\gamma}_{n-\alpha}\right) - 2|\vec{s}|^2\right)\\
    &= \sum_{\alpha=1-d}^d r_\alpha(A_\alpha + G_\alpha) + L(r_0-|\vec{s}|^2)
\end{split}
\end{align}

\noindent for some constants $\{r_\alpha(\vec{s})\}$. These are real functions of $\vec{s}$. While we do not need the explicit form of these coefficients, the computation can be done algorithmically. Algorithm \ref{alg:compute-talpha} in Appendix \ref{appendix:algorithm-t} gives these coefficients up to an overall rescaling. 

We may wonder if there has been any cancellation of coefficients in this representation that would give a non-vanishing contribution in general. However, for any linear dependence (aside from $G_k = G_{-k}$, which is true for all representations) to arise, we would need the transformation \cref{eqn:Htilde-definition-rotations} to produce generators $A_l, G_m$ with $l,m $ of order $ L$. However, \cref{eqn:Htilde-Majoranas} is true for all $L$. If we consider $L \gg d,p$, then \cref{eqn:Htilde-definition-rotations} cannot produce sufficiently large $l,m$. Therefore, $r_\alpha = a_\alpha = g_\alpha$ is the only possible set of $L$-independent coefficients generated by \cref{eqn:Htilde-definition-rotations} for $N=2$. Moreover, \cref{eqn:generalHtilde=aA+gG} contains no constant term, and no linear combination of $A_l, G_m$ can give rise to this. Consequently, Algorithm \ref{alg:compute-talpha} must give $r_0 = |\vec{s}|^2$, and
\begin{align} \label{eqn:Htilde-final-general-N}
    \tilde{H}_A=\sum_{\alpha=1-d}^d r_\alpha(A_\alpha + G_\alpha)
\end{align}
\noindent holds for all $N$. Note that the $\{r_\alpha(\vec{s})\}$ are independent of $N$ and $L$.

As noted in Section \ref{subsec:chiral-clock-intro}, $\ket{\psi_0}$ is an eigenstate of $(A_\alpha+G_\alpha)$. Thus, it is an eigenstate of $\tilde{H}_A$. Reversing the transformation \cref{eqn:Htilde-definition-rotations}, we see that $\ket{\varphi}= \mathrm{M}^{(d)}\dots \mathrm{M}^{(1)}\ket{\psi_0}$ must therefore be an eigenstate of $H_A$, provided that $|b_k| \neq 1$ for all $k$. Following a similar procedure to that presented in Ref. \cite{Jones21}, this can be expressed in MPS form: each operator $\mathrm{M}^{(k)}$ is a local operator that can be expressed as an MPO with finite bond dimension, and $\ket{\psi_0}$ is a product state. Acting with a series of finite bond dimension MPOs on a product state gives an exact MPS.

Thus far, we have restricted ourselves to Hamiltonians with $f(z) = g(z)^2$. However, the Hamiltonians $f(z) = -g(z)^2$ take each $A_k \mapsto -A_k$. If $\{A_l, G_m|m,l\in\mathbb{Z}\}$ obey the Onsager algebra, then the generators $\{-A_l, G_m|m,l\in\mathbb{Z}\}$ obey the same structure. Therefore, to obtain the ground state of $f(z) = -g(z)^2$, we merely replace each $A_k\mapsto -A_k$ in every $\mathrm{M^{(k)}}$, and replace the ground state $\ket{\psi_p}$ of $A_p$ with the ground state $\ket{\psi_p^-}$ of $-A_p$.

Generalising this proof to any even $p=2k$ is trivial: the mapping $A_k \mapsto A_{k+1}, G_k \mapsto G_k$ preserves the structure of the Onsager algebra and \cref{eqn:unitary-pivot} ensures that the ground state $\ket{\psi_{2k}}$ is unique. However, extending to any odd $p=2k+1$ requires more care, as Kramers-Wannier duality tells us that the ground state of $A_1$ is $N$-fold degenerate. Nevertheless, \cref{result:theorem-1} holds for all $p$; $A_\alpha +G_\alpha$ commutes with $r$, and is therefore diagonalised for all $ \alpha$  by the $N$ eigenstates of $r$ with distinct charges $0\leq Q\leq N-1 $. We give a more detailed discussion in Appendix \ref{appendix:kramers-wannier}.

\subsection{Proof of Result 2} \label{subsec:theorem-2-proof}
We have established that, provided $|b_k| \neq 1$ for all $k$, the state $\ket{\varphi}$ given in \cref{eqn:gs-conjecture-varphi} is an eigenstate of the Hamiltonian $H_A$ with $f(z)=\pm z^p g(z)^2$. Results of perturbation theory in finite-dimensional Hilbert spaces from Ref. \cite{Kato66} demonstrate that, provided the gap does not close, the ground state is an analytic function of $\vec{s}$. We apply this below to show that $\ket{\varphi(\vec{s})}$ is the ground state provided $H_A\in\mathcal{S}$.
\subsubsection{The ground state in the gapped region containing \texorpdfstring{$A_0$}{A_0}}
We begin by noting that $\ket{\varphi(\vec{S})} = \ket{\psi_0}$ and $H_A(\vec{S}) = A_0$ for the vector $\vec{S} = (1,0,0,\dots)$. Thus, $\ket{\varphi(\vec{s})}$ is the ground state of $H_A(\vec{s})$ at $\vec{s} = \vec{S}$. Provided that the gap does not close, the ground state must be an analytic function of $\vec{s}$. Therefore, since $\ket{\varphi(\vec{s})}$ is an analytic function of $\vec{s}$ containing the ground state at $\vec{s}=\vec{S}$, it is the ground state everywhere in the gapped region surrounding $A_0$.

Note that when the gap closes, we find degeneracies, and this argument breaks down. We have therefore proved only that $\ket{\varphi(\vec{s})}$ is the ground state of $H_A(\vec{s})$ in the gapped region containing $A_0$.

\subsubsection{The ground state in the gapped region containing \texorpdfstring{$A_l$}{A_l}}

The gapped region containing $A_0$ has every $|b_k| < 1$. If we instead have some $|b_k| > 1$, then we can use the identity 

\begin{align}
    \text{arctanh}(b_k) = \text{arctanh}\left(\frac{1}{b_k}\right) + \frac{i \pi}{2}
\end{align}

\noindent to write 

\begin{align}
    \mathrm{M}^{(k)} = \exp\left(-\beta_kA_k\right) = \exp\left(-\eta_kA_k\right)U_k\ ,
\end{align}

\noindent where we have defined 

\begin{align}
    \eta_k = 2\text{ arctanh}\left(\frac{1}{b_k}\right) \quad\quad\quad  U_k = \exp\left(-i\pi A_k\right)\ .
\end{align}

\noindent We can use the relation $U_k e^{-\beta_q A_q}  = e^{-\beta_q A_{2k-q}}U_k$ from \cref{eqn:pivot} to move all the unitary operators $U_k$ so that they act on the initial state $\ket{\psi_0}$. As shown in Ref. \cite{Jones24}, this gives the ground state $\ket{\psi_l}$ of another fixed-point Hamiltonian $A_l$. Therefore, we are always able to write

\begin{align}
    \ket{\varphi} = \tilde{\mathrm{M}}_{n_d}^{(d)}\dots \tilde{\mathrm{M}}_{n_1}^{(1)} \ket{\psi_l}\ ,
\end{align}

\noindent where $l \in \mathbb{Z}$ and 

\begin{align}
    \tilde{\mathrm{M}}_{n}^{(k)} = \exp\left(-2\text{ arctanh}(\tilde{b}_k) A_{n}\right)
\end{align}

\noindent for $|\tilde{b}_k|<1$ and some integer $n$.

Although the state $\ket{\varphi(\vec{s})}$ is not smoothly connected to $\ket{\psi_0}$ along a path of gapped Hamiltonians $H_A$ if any of the $|b_k|>1$, it is always smoothly connected to a ground state $\ket{\psi_l}$ of a Hamiltonian $A_l$ for some $l \in \mathbb{Z}$. At this point, $H_A(\vec{s}) = A_l$. Therefore, the eigenstate $\ket{\varphi(\vec{s})}$ is the ground state provided $ H_A \in \mathcal{S}$.

\subsection{Proof of Result 3} \label{subsec:theorem-3-proof}
We can write the Laurent polynomial \cref{eqn:laurent-polynomial-define} describing a general Hamiltonian $H_A$ in the family \cref{eqn:general-chiral-clock-family} as 

\begin{align}
\begin{split}
    f(z) &= \sigma z^{P_0} \prod_{j=1}^{P_z} (z-z_j)\prod_{k=1}^{P_\zeta}(z-\zeta_k)\\
    &= \sum_{r=0}^{P_z+P_\zeta} f_r z^{r+P_0}\ ,
\end{split}
\end{align}

\noindent where $\sigma \in \mathbb{R}$, $P_0 \in \mathbb{Z}$, $\{P_\zeta, P_z\}$ are non-negative integers, $|z_j|<1$, and $|\zeta_k|>1$. Note that this polynomial has no zeros on the unit circle. Such zeros correspond to a gap closing in $\mathcal{S}$ and, as discussed in Appendix \ref{appendix:schur-cohn-algorithm}, the existence of zeros on the unit circle will give a $|b_k|=1$, which we exclude. If we now define 

\begin{align} \label{eqn:g(z)-exact}
    g(z) = \prod_{j=1}^{P_z} \sqrt{1-\frac{z_j}{z}}\prod_{k=1}^{P_\zeta}\sqrt{1-\frac{z}{\zeta_k}}\ ,
\end{align}

\noindent then, for $q = P_0 +P_z$, we have $f(z) =\pm z^q g(z)^2$. As overall scale is unimportant, we normalise the constant coefficient to $\pm1$.

This polynomial $g(z)$ can be expanded as a Laurent series

\begin{align}
    g(z) = \sum_{n=-\infty}^\infty s_n z^n\ ,
\end{align}

\noindent which is well-defined in the annulus $\mathscr{A}$ given by

\begin{align}
    \text{max}_j\{|z_j|\} < |z| < \text{min}_k\{|\zeta_k|\}\ .
\end{align}

\noindent Now, truncating this series, we obtain 

\begin{align}
    g_D(z) = \sum_{n=-D}^D s_n z^n\ ,
\end{align}

\noindent and 

\begin{align}
\begin{split}
    f_D(z) = \pm\sum_{n=-D}^D \sum_{m=-D}^D s_n s_m z^{n+m+q}=\pm\sum_{r=-2D}^{2D} c_r z^{r+q}\ ,
\end{split}
\end{align}

\noindent which defines the Hamiltonian $H_A^{(D)}$. From this, we can define the new Hamiltonian $\delta H_A^{(D)}  = H_A^{(D)} - H_A^{}$ with Laurent polynomial

\begin{align}\label{eqn:deltaf}
    \delta f_D(z) = \pm\sum_{r=-2D}^{2D} (c_r-f_{r+P_z})z^{r+q}\ .
\end{align}

\noindent In any sector, consider the ground states $\ket{\varphi_D}$ and $\ket{\varphi}$ of $H_A^{(D)}$ and $H_A$, respectively. Define the energy densities 

\begin{equation}
    \mathcal{E}_A = \frac{1}{L}\bra{\varphi}H_A\ket{\varphi} \quad\quad\quad\quad \mathcal{E}_A^{(D)} = \frac{1}{L}\bra{\varphi_D}H_A\ket{\varphi_D} .
\end{equation}

\noindent From variational principles, we have 

\begin{align}
    0\leq \mathcal{E}_A^{(D)} - \mathcal{E}_A^{} \leq \frac{1}{L}\bra{\varphi}\delta H_A^{(D)}\ket{\varphi} - \frac{1}{L} \bra{\varphi_D}\delta H_A^{(D)}\ket{\varphi_D}.
\end{align}

\noindent By definition of the spectral norm, for any normalised state $\ket{\psi}$, we have $|\bra{\psi}\delta H_A^{(D)}\ket{\psi}| \leq ||\delta H_A^{(D)}||$. Therefore, using the triangle inequality, the energy densities differ by

\begin{align}
\begin{split}
    |\mathcal{E}_A^{(D)} - \mathcal{E}_A^{}| &\leq \frac{2||\delta H_A^{(D)}||}{L} \\
    &\leq \frac{2}{L}\sum_{r=-2D}^{2D} |c_r-f_{r+P_z}| \,\, ||A_{r+q}||\\
    &\leq 2\mathscr{E}_\text{max} \sum_{r=-2D}^{2D} |c_r-f_{r+P_z}|\ ,
\end{split}
\end{align}

\noindent where $\mathscr{E}_{\text{max}}$ is a bound on the energy density of the Onsager generators $A_l$ (assumed in general to be $O(1)$). In the chiral clock model,
$\mathscr{E}_{\text{max}} = (N-1)/2N$ \cite{Jones24}.

The coefficients in \cref{eqn:deltaf} can be evaluated using residue calculus. This allows us to bound the magnitude of each term as follows:

\begin{align}
\begin{split}
    |c_r-f_{r+P_z}| &= \left|\frac{1}{2\pi i} \oint_{|z|=\rho}   \frac{\delta f_D(z)}{z^{r+q+1}}\,\mathrm{d}z \right|\\
    &\leq \frac{M_\rho(D)}{\rho^{r+q}}\ ,
\end{split}
\end{align}

\noindent where $\text{max}_j\{|z_j|\} < \rho < \text{min}_k\{|\zeta_k|\}$ and

\begin{align}
    M_\rho(D) = \sup_{|z|=\rho} |\delta f_D(z)|\ .
\end{align}

\noindent Choosing $\text{max}_j\{|z_j|\} < \rho_-<1<\rho_+ < \text{min}_k\{|\zeta_k|\}$, we can write

\begin{align}
\begin{split}
    \sum_{r=-2D}^{2D} |c_r-f_{r+P_z}| &\leq M_{\rho_+}(D) \sum_{r=-q}^{2D} \frac{1}{\rho_+^{r+q}}  + M_{\rho_-}(D) \sum_{r=-2D}^{-1-q} \frac{1}{\rho_-^{r+q}}\\
    &\leq  \text{max}\{M_{\rho_+}(D), M_{\rho_-}(D)\}\Sigma\ ,
\end{split}
\end{align}

\noindent where we have defined the finite

\begin{align}
    \Sigma = \left(\frac{\rho_+}{\rho_+ -1}+\frac{\rho_-}{1-\rho_-}\right)\ .
\end{align}

\noindent However, convergence of $f_D(z)$ to $f(z)$ in the annulus $\mathscr{A}$ implies that for any $\epsilon>0$ we can fix a sufficiently large $D$ (and in the background an $L\gg D$) such that $M_{\rho_\pm}(D) < \epsilon$, giving

\begin{align}
    \sum_{r=-2D}^{2D} |c_r-f_{r+P_z}|\leq \epsilon \Sigma\ ,
\end{align}

\noindent and therefore that 

\begin{align}
    |\mathcal{E}_A^{(D)} - \mathcal{E}_A^{}| <  2\epsilon \Sigma \mathscr{E}_{\text{max}}\ .
\end{align}

\noindent Hence, the MPS skeleton is dense: for any Hamiltonian $H_A\in \mathcal{S}$ in the Onsager-integrable chiral clock class, we can construct a Hamiltonian $H_A^{(D)}$ on the skeleton with a ground state of the form

\begin{align}
    \ket{\varphi_D} = \mathrm{M}^{(d)}\dots \mathrm{M}^{(1)} \ket{\psi_p^\pm} \ ,
\end{align}

\noindent with an energy density with respect to $H_A$ approaching the ground-state energy density of $H_A$. Thus, we have a method for approximating the ground state of any Hamiltonian in $\mathcal{S}$. Moreover, by the same reasoning, the ground state of $H_A^{(D)}$ in any particular sector approximates the ground state of $H_A$ in that sector for sufficiently large $D$.

\subsection{Proof of Result 4} \label{subsec:theorem-4-proof}

As shown in Section \ref{subsec:theorem-1-proof}, we can write 

\begin{align}
    \tilde{H}_A = \sum_{\alpha=1-d}^dr_\alpha(A_\alpha+G_\alpha)\ .
\end{align}

\noindent As noted in  \Cref{subsec:chiral-clock-intro}, the states $\ket{\phi^{(P)}_{E_0+1/N,\,p=0}}$ are eigenstates of $(A_\alpha+G_\alpha)$. Thus, these states are eigenstates of $\tilde{H}_A$. Reversing the transformation \cref{eqn:Htilde-definition-rotations}, we see that $\ket{\chi_{p=0}^{(P)}}= \mathrm{M}^{(d)}\dots \mathrm{M}^{(1)}\ket{\phi^{(P)}_{E_0+1/N,\, p=0}}$ are eigenstates of $H_A$, provided that $|b_k| \neq 1$ for all $k$. 

Similarly to the ground state, $\ket{\phi^{(P)}_{E_0+1/N,\, p=0}}$ has an exact MPS representation. In analogy with the $W$-state \cite{Cirac21}, the most efficient MPS representation is

\begin{align}
\begin{split}
    \ket{\phi^{(P)}_{E_0+1/N,\, p=0}} &= \frac{1}{\sqrt{L}}\sum_{j=1}^L e^{-iPj}Z_j^\dag \ket{\psi_0}\\ 
    &= \sum_{\{\sigma_j \in \{0,1\}\}} \bra{L} \mathcal{A}_1^{(\sigma_1)} \dots \mathcal{A}_L^{(\sigma_L)}\ket{R} \ket{\sigma_1 \dots \sigma_L}\ , 
\end{split}
\end{align}

\noindent where we have defined the vectors

\begin{align}
    \ket{L} = \begin{pmatrix}1\\0\end{pmatrix} \quad\quad\quad \ket{R} = \begin{pmatrix}0\\1\end{pmatrix}
\end{align}

\noindent and the site-dependent matrices

\begin{align}
    \mathcal{A}_j^{(0)} = \begin{pmatrix}1 & 0\\0 & 1\end{pmatrix} \quad\quad\quad \mathcal{A}_j^{(1)} = \frac{1}{\sqrt{L}}\begin{pmatrix}0 & e^{- i P j}\\0 & 0\end{pmatrix}\ .
\end{align}

\noindent As discussed in Section \ref{subsec:theorem-1-proof}, each $\mathrm{M}^{(k)}$ can be expressed as an MPO with finite bond dimension. Applying a series of such MPOs to an exact MPS yields another MPS of finite bond dimension. Thus, the eigenstate $\ket{\chi_{p=0}^{(P)}}$ has a representation as an exact MPS. 

Moreover, for $p\neq 0$, keeping $p$ even, we can use \cref{eqn:unitary-pivot} to obtain eigenstates \begin{align}\ket{\chi_{p}^{(P)}}= \mathrm{M}^{(d)}\dots \mathrm{M}^{(1)}\ket{\phi^{(P)}_{E_0+1/N,\, p}}\end{align} as defined in \cref{eqn:excited-state-conjecture}. Note that the pivot procedure defined in Ref. \cite{Jones24} ensures that $\ket{\phi^{(P)}_{E_0+1/N,\, p}}$ also has an exact MPS representation. Unlike in the ground state case, it is unclear whether this could be generalised for odd $p$, as Kramers-Wannier dual Hamiltonians with periodic boundary conditions cannot have a single domain wall.

\subsection{Constraints on coefficients \texorpdfstring{$\{r_\alpha\}$}{r_α}} \label{subsec:constraints-on-ralpha}
In this section, we identify some constraints on the coefficients $r_\alpha$ appearing above.
We have 

\begin{align}
    \tilde{H}_A \ket{\psi_0} = \varepsilon_\varphi \ket{\psi_0} \quad\implies\quad H_A\ket{\varphi} = \varepsilon_\varphi \ket{\varphi}.
\end{align}

\noindent Hence, the energy of the eigenstate $\ket{\varphi}$ of the Hamiltonian $H_A$ with $f(z)=g(z)^2$ is given by

\begin{align}\label{eqn:gs-energy-from-eigenstate}
\begin{split}
    \varepsilon_\varphi &= \bra{\varphi}H_A\ket{\varphi}= \bra{\psi_0}\tilde{H}_A\ket{\psi_0}\\
    &= \sum_{\alpha=1-d}^d r_\alpha\bra{\psi_0}A_\alpha \ket{\psi_0}\\
    &= r_0 \bra{\psi_0}A_0\ket{\psi_0} +\sum_{\alpha=1}^d (r_\alpha+r_{-\alpha})\bra{\psi_0}A_\alpha \ket{\psi_0}\ ,
\end{split}
\end{align}

\noindent as the ladder operators \cref{eqn:ladder-operators} impose that $\bra{\psi}A_\alpha\ket{\psi} = \bra{\psi}A_{-\alpha}\ket{\psi}$ for any eigenstate $\ket{\psi}$ of $A_0$.

Defining $E^\pm_k = J_x^k \pm i J_y^k$, it follows from \cref{eqn:su2-transformation} that

\begin{align}
    \bra{\psi_0}A_\alpha\ket{\psi_0} = \sum_{k=1}^n \cos(m\theta_k) \bra{\psi_0}J_k^x\ket{\psi_0} = -S \sum_{k=1}^n \cos(m\theta_k)
\end{align}

\noindent for spin $S$. For these Hamiltonians $H_A$, we can use this and \cref{eqn:spectrum-form} to write the ground state energy as

\begin{align} \label{eqn:gs-energy-baxter}
\begin{split}
    \varepsilon_0 = -S \sum_{l,m=0}^d s_l s_m \sum_{k=1}^n  \cos((l-m)\theta_k) = \sum_{l,m=0}^d s_l s_m \bra{\psi_0}A_{l-m}\ket{\psi_0}.
\end{split}
\end{align}

We know that $\ket{\varphi}$ is the ground state of the ``ground-state Onsager sector'' for Hamiltonians $H_A$. Thus, $\varepsilon_\varphi = \varepsilon_0$. From this, we can conclude that $r_0 = |\vec{s}|^2$ (as we had already argued) and that 

\begin{align}
    r_\alpha + r_{-\alpha} = 2\sum_{\substack{l,m=0\\l-m=\alpha}}^d s_l s_m \quad\quad\quad \alpha \in \{1,2,\dots,d\}\ .
\end{align}

\noindent Algorithm \ref{alg:compute-talpha} will therefore give coefficients satisfying this relationship.
\section{\label{sec:outlook}Outlook}
In this work, we uncovered dense MPS skeletons in general classes of Onsager-integrable Hamiltonians, each analogous to that in Ref. \cite{Jones21}. While certain results hold for any representation, we deal primarily with the Onsager-integrable chiral clock chains---these are the key physical representations of this class. Due to the interacting nature of these models, any analysis is more involved than in the free-fermion classes considered previously. Nevertheless, we constructed the ground state explicitly for chiral clock Hamiltonians lying on the skeleton in gapped regions of the phase diagram containing any fixed-point Hamiltonian $A_l$. Moreover, we showed that the constructed state remains an eigenstate everywhere on the skeleton, excluding a measure-zero set of cases. Given that the skeleton is dense, we also have a method for approximating the ground state of any model in these gapped regions as the limit of a sequence of states of the form \cref{eqn:gs-conjecture-varphi}.

To prove the existence of the MPS skeleton, we worked in a specific representation of the Onsager algebra, considering Hamiltonians of the form $H_A = \pm\sum_{k,q} s_k s_q A_{k+q+p}$. We emphasise that, even fixing $N=2$, the method used differs from the proof presented in Ref. \cite{Jones21}. However, it is not fully independent, since we utilise features of the frustration-free form of the Hamiltonian derived in this work. We note that the methods used in earlier work on the BDI skeleton rely heavily on the free-fermion representation. It would be interesting to explore whether our models are frustration-free for $N>2$, potentially using the parafermionic representation, and to understand whether the Witten conjugation method can be used in this context \cite{Wouters21}. Note that any such application will be complicated by the level crossing transitions at the boundary of $\mathcal{S}$.

We calculated the ground state expectation of the disorder operator for the $d=1$ leg of the skeleton, showing how the well-understood $N=2$ formula generalises to all even $N$. Whether the same generalisation occurs for larger $d$ remains an interesting question that we plan to explore in future work. Our derivation of the disorder operator on the skeleton gives a new route to understanding correlations in this family of models. Away from the superintegrable chiral Potts line, a formula for the disorder parameter is not known, despite the analogous result being established for the $N=2$ case using Toeplitz determinant theory \cite{Jones19}. Baxter gives a determinantal formula for this operator on the superintegrable chiral Potts line \cite{Baxter10}. We expect that it will be possible to construct similar determinantal results away from this line using the MPS skeleton: if we establish a closed formula on the skeleton, we could use the density of the skeleton to obtain a general formula. 

Importantly, we contrast our results with existing literature on MPS in Bethe ansatz solvable models, as Bethe ansatz eigenstates can be placed in an MPS framework \cite{Alcaraz03,Katsura10,Murg12,Ruiz24,Ruiz25}. While this gives an MPS form for all eigenstates, the corresponding bond dimension can diverge in the thermodynamic limit (dependent on the magnetisation of the state), and naturally relies on a $U(1)$ symmetry labelling this magnetisation. In $H_A$, the charge is conserved modulo $N$. Thus, we cannot apply these results to find exact MPS ground states of our family of models. It would be interesting to see if we could use our exact ground states to feed into a further Bethe ansatz analysis, or to connect to the eigenvector constructions of Refs.~\cite{AuYang08,AuYang09}. Our MPS skeleton states could also be useful initial states for numerics.

Our MPS construction gives an upper bound on the number of non-zero entanglement eigenvalues of the ground state. In Refs.~\cite{Jones21,Jones23}, this construction was conjectured to be optimal for $N=2$, which could be proved in certain cases. It is unclear whether this optimality also applies to the clock models.

In Ref.~\cite{Shibata20}, quantum scar states are constructed via imaginary time evolutions with certain Onsager algebra elements. These are MPS eigenstates in a non-integrable variant of the model in Ref.~\cite{Vernier19}. It would be interesting to understand if our construction would be useful in this context.

Finally, we constructed a further set of $L$ exact MPS eigenstates. These are excited states in the aforementioned gapped regions of the phase diagram. We could use these states to place partial constraints on the location of the phase boundaries: the gap will certainly close when any one of these states becomes degenerate with $\ket{\varphi}$. However, this would not determine the phase boundaries entirely. Therefore, we leave the characterisation of the phase boundaries for further work. 

\section*{Acknowledgements}
We are grateful to Paul Fendley and Riccardo Senese for illuminating discussions.
We thank Abhishodh Prakash, Ruben Verresen and Harvey Weinberger for insightful discussions and for comments on the manuscript. IC gratefully acknowledges support from EPSRC.

\bibliography{arxiv.bbl}

\providecommand{\href}[2]{#2}\begingroup\raggedright\begin{thebibliography}{10}

\bibitem{Korepin97}
V.E.~Korepin, N.~Bogoliubov and A.~Izergin, \emph{Quantum inverse scattering method and correlation functions}, vol.~3, Cambridge University Press (1997).

\bibitem{Baxter16}
R.J.~Baxter, \emph{Exactly solved models in statistical mechanics}, Elsevier (2016).

\bibitem{White92}
S.R.~White, \emph{Density matrix formulation for quantum renormalization groups}, \href{https://doi.org/10.1103/PhysRevLett.69.2863}{\emph{Phys. Rev. Lett.} {\bfseries 69} (1992) 2863}.

\bibitem{Schollwock11}
U.~Schollw\"{o}ck, \emph{The density-matrix renormalization group in the age of matrix product states}, \href{https://doi.org/10.1016/j.aop.2010.09.012}{\emph{Annals of Physics} {\bfseries 326} (2011) 96–192}.

\bibitem{Hastings07}
M.B.~Hastings, \emph{An area law for one-dimensional quantum systems}, \href{https://doi.org/10.1088/1742-5468/2007/08/p08024}{\emph{Journal of Statistical Mechanics: Theory and Experiment} {\bfseries 2007} (2007) P08024–P08024}.

\bibitem{Cirac21}
J.I.~Cirac, D.~P\'erez-Garc\'{\i}a, N.~Schuch and F.~Verstraete, \emph{Matrix product states and projected entangled pair states: Concepts, symmetries, theorems}, \href{https://doi.org/10.1103/RevModPhys.93.045003}{\emph{Rev. Mod. Phys.} {\bfseries 93} (2021) 045003}.

\bibitem{Affleck87}
I.~Affleck, T.~Kennedy, E.H.~Lieb and H.~Tasaki, \emph{{Rigorous results on valence-bond ground states in antiferromagnets}}, \href{https://doi.org/10.1103/PhysRevLett.59.799}{\emph{Phys. Rev. Lett.} {\bfseries 59} (1987) 799}.

\bibitem{Pollmann10}
F.~Pollmann, A.M.~Turner, E.~Berg and M.~Oshikawa, \emph{Entanglement spectrum of a topological phase in one dimension}, \href{https://doi.org/10.1103/PhysRevB.81.064439}{\emph{Phys. Rev. B} {\bfseries 81} (2010) 064439}.

\bibitem{Chen11}
X.~Chen, Z.-C.~Gu and X.-G.~Wen, \emph{Classification of gapped symmetric phases in one-dimensional spin systems}, \href{https://doi.org/10.1103/PhysRevB.83.035107}{\emph{Phys. Rev. B} {\bfseries 83} (2011) 035107}.

\bibitem{Schuch11}
N.~Schuch, D.~P\'erez-Garc\'{\i}a and I.~Cirac, \emph{Classifying quantum phases using matrix product states and projected entangled pair states}, \href{https://doi.org/10.1103/PhysRevB.84.165139}{\emph{Phys. Rev. B} {\bfseries 84} (2011) 165139}.

\bibitem{Turner11}
A.M.~Turner, F.~Pollmann and E.~Berg, \emph{Topological phases of one-dimensional fermions: An entanglement point of view}, \href{https://doi.org/10.1103/PhysRevB.83.075102}{\emph{Phys. Rev. B} {\bfseries 83} (2011) 075102}.

\bibitem{Barouch71}
E.~Barouch and B.M.~McCoy, \emph{{Statistical Mechanics of the $XY$ Model. II. Spin-Correlation Functions}}, \href{https://doi.org/10.1103/PhysRevA.3.786}{\emph{Phys. Rev. A} {\bfseries 3} (1971) 786}.

\bibitem{Kurmann82}
J.~Kurmann, H.~Thomas and G.~M{\"u}ller, \emph{Antiferromagnetic long-range order in the anisotropic quantum spin chain}, {\emph{Physica A: Statistical Mechanics and its Applications} {\bfseries 112} (1982) 235}.

\bibitem{Muller85}
G.~M\"uller and R.E.~Shrock, \emph{{Implications of direct-product ground states in the one-dimensional quantum XYZ and XY spin chains}}, \href{https://doi.org/10.1103/PhysRevB.32.5845}{\emph{Phys. Rev. B} {\bfseries 32} (1985) 5845}.

\bibitem{Chung01}
M.-C.~Chung and I.~Peschel, \emph{Density-matrix spectra of solvable fermionic systems}, \href{https://doi.org/10.1103/PhysRevB.64.064412}{\emph{Phys. Rev. B} {\bfseries 64} (2001) 064412}.

\bibitem{Franchini07}
F.~Franchini, A.R.~Its and V.E.~Korepin, \emph{{Renyi entropy of the XY spin chain}}, \href{https://doi.org/10.1088/1751-8113/41/2/025302}{\emph{Journal of Physics A: Mathematical and Theoretical} {\bfseries 41} (2007) 025302}.

\bibitem{Wolf06}
M.M.~Wolf, G.~Ortiz, F.~Verstraete and J.I.~Cirac, \emph{{Quantum Phase Transitions in Matrix Product Systems}}, \href{https://doi.org/10.1103/PhysRevLett.97.110403}{\emph{Phys. Rev. Lett.} {\bfseries 97} (2006) 110403}.

\bibitem{Smith22}
A.~Smith, B.~Jobst, A.G.~Green and F.~Pollmann, \emph{{Crossing a topological phase transition with a quantum computer}}, \href{https://doi.org/10.1103/PhysRevResearch.4.L022020}{\emph{Phys. Rev. Res.} {\bfseries 4} (2022) L022020}.

\bibitem{Jones21}
N.G.~Jones, J.~Bibo, B.~Jobst, F.~Pollmann, A.~Smith and R.~Verresen, \emph{{Skeleton of matrix-product-state-solvable models connecting topological phases of matter}}, \href{https://doi.org/10.1103/physrevresearch.3.033265}{\emph{Physical Review Research} {\bfseries 3} (2021) }.

\bibitem{Jones23}
N.G.~Jones and R.~Verresen, \emph{Exact correlations in topological quantum chains}, \href{https://doi.org/10.3842/sigma.2023.098}{\emph{Symmetry, Integrability and Geometry: Methods and Applications} (2023) }.

\bibitem{Altland97}
A.~Altland and M.R.~Zirnbauer, \emph{Nonstandard symmetry classes in mesoscopic normal-superconducting hybrid structures}, \href{https://doi.org/10.1103/PhysRevB.55.1142}{\emph{Phys. Rev. B} {\bfseries 55} (1997) 1142}.

\bibitem{Lieb61}
E.~Lieb, T.~Schultz and D.~Mattis, \emph{Two soluble models of an antiferromagnetic chain}, {\emph{Annals of Physics} {\bfseries 16} (1961) 407}.

\bibitem{Franchini11}
F.~Franchini, A.~Its, V.E.~Korepin and L.A.~Takhtajan, \emph{{Spectrum of the density matrix of a large block of spins of the XY model in one dimension}}, {\emph{Quantum Information Processing} {\bfseries 10} (2011) 325}.

\bibitem{Onsager44}
L.~Onsager, \emph{{Crystal statistics. 1. A Two-dimensional model with an order disorder transition}}, \href{https://doi.org/10.1103/PhysRev.65.117}{\emph{Phys. Rev.} {\bfseries 65} (1944) 117}.

\bibitem{Dolan82}
L.~Dolan and M.~Grady, \emph{{Conserved charges from self-duality}}, \href{https://doi.org/10.1103/PhysRevD.25.1587}{\emph{Phys. Rev. D} {\bfseries 25} (1982) 1587}.

\bibitem{Ahn91}
C.~Ahn and K.~Shigemoto, \emph{{Onsager algebra and integrable lattice models}}, \href{https://doi.org/10.1142/S021773239100405X}{\emph{Mod. Phys. Lett. A} {\bfseries 6} (1991) 3509}.

\bibitem{vonGehlen85}
G.~{von Gehlen} and V.~Rittenberg, \emph{{$Z_n$-symmetric quantum chains with an infinite set of conserved charges and $Z_n$ zero modes}}, \href{https://doi.org/https://doi.org/10.1016/0550-3213(85)90350-5}{\emph{Nuclear Physics B} {\bfseries 257} (1985) 351}.

\bibitem{Baxter88}
R.J.~Baxter, \emph{{The superintegrable chiral Potts model}}, \href{https://doi.org/10.1016/0375-9601(88)91014-6}{\emph{Phys. Lett. A} {\bfseries 133} (1988) 185}.

\bibitem{Baxter89}
R.J.~Baxter, \emph{{Superintegrable chiral Potts model: Thermodynamic properties, an inverse model and a simple associated Hamiltonian}}, \href{https://doi.org/10.1007/BF01023632}{\emph{J. Statist. Phys.} {\bfseries 57} (1989) 1}.

\bibitem{Davies90}
B.~Davies, \emph{{Onsager's algebra and superintegrability}}, \href{https://doi.org/10.1088/0305-4470/23/12/010}{\emph{Journal of Physics A: Mathematical and General} {\bfseries 23} (1990) 2245}.

\bibitem{Jones24}
N.G.~Jones, A.~Prakash and P.~Fendley, \emph{Pivoting through the chiral-clock family}, \href{https://doi.org/10.21468/scipostphys.18.3.094}{\emph{SciPost Physics} {\bfseries 18} (2025) }.

\bibitem{Jones25}
N.G.~Jones, R.~Thorngren, R.~Verresen and A.~Prakash, \emph{{Charge pumps, pivot Hamiltonians, and symmetry-protected topological phases}}, \href{https://doi.org/10.1103/rtq1-pplf}{\emph{Phys. Rev. B} {\bfseries 112} (2025) 165123}.

\bibitem{Vernier19}
E.~Vernier, E.~O’Brien and P.~Fendley, \emph{{Onsager symmetries in $U(1)$-invariant clock models}}, \href{https://doi.org/10.1088/1742-5468/ab11c0}{\emph{Journal of Statistical Mechanics: Theory and Experiment} {\bfseries 2019} (2019) 043107}.

\bibitem{Gioia25}
L.~Gioia and R.~Thorngren, \emph{Exact chiral symmetries of $3+1\mathrm{D}$ hamiltonian lattice fermions}, \href{https://doi.org/10.1103/s5q2-317m}{\emph{Phys. Rev. Lett.} {\bfseries 136} (2026) 061601}.

\bibitem{Pace25}
S.D.~Pace, A.~Chatterjee and S.-H.~Shao, \emph{{Lattice T-duality from non-invertible symmetries in quantum spin chains}}, \href{https://doi.org/10.21468/SciPostPhys.18.4.121}{\emph{SciPost Phys.} {\bfseries 18} (2025) 121}.

\bibitem{Su25}
L.~Su, \emph{{$\mathbb{Z}_{2}$ gauging and self-dualities of the $XX$ model and its cousins}}, \href{https://doi.org/10.1103/c16r-fdfb}{\emph{Phys. Rev. B} {\bfseries 112} (2025) 035105}.

\bibitem{Roan11}
S.-S.~Roan, \emph{{Eigenvectors of an Arbitrary Onsager Sector in Superintegrable $\tau^{(2)}$-model and Chiral Potts Model}},  \href{https://arxiv.org/abs/1003.3621}{{\ttfamily 1003.3621}}.

\bibitem{McCoy91}
B.M.~McCoy, \emph{{The chiral Potts model: From physics to mathematics and back}},  in \emph{{ICM-90 Satellite Conference Proceedings: Special Functions}}, pp.~245--259, Springer, 1991.

\bibitem{VonGehlen01}
G.~Von~Gehlen and S.-S.~Roan, \emph{{The Superintegrable Chiral Potts Quantum Chain and Generalized Chebyshev Polynomials}},  in \emph{{Integrable Structures of Exactly Solvable Two-Dimensional Models of Quantum Field Theory}}, S.~Pakuliak and G.~von Gehlen, eds., (Dordrecht), pp.~155--172, Springer Netherlands (2001), \href{https://doi.org/10.1007/978-94-010-0670-5_10}{DOI}.

\bibitem{Albertini89b}
G.~Albertini, B.M.~McCoy and J.H.H.~Perk, \emph{{Eigenvalue spectrum of the superintegrable chiral Potts model}},  in \emph{{Advanced Studies in Pure Mathematics}}, vol.~19, pp.~1--55, 1989.

\bibitem{Dasmahapatra93}
S.~Dasmahapatra, R.~Kedem and B.M.~McCoy, \emph{{Spectrum and completeness of the three state superintegrable chiral Potts model}}, \href{https://doi.org/10.1016/0550-3213(93)90662-9}{\emph{Nucl. Phys. B} {\bfseries 396} (1993) 506} [\href{https://arxiv.org/abs/hep-th/9204003}{{\ttfamily hep-th/9204003}}].

\bibitem{Iorgov10}
N.~Iorgov, S.~Pakuliak, V.~Shadura, Y.~Tykhyy and G.~von Gehlen, \emph{Spin operator matrix elements in the superintegrable chiral potts quantum chain}, \href{https://doi.org/10.1007/s10955-010-9972-1}{\emph{Journal of Statistical Physics} {\bfseries 139} (2010) 743–768}.

\bibitem{AuYang08}
H.~Au-Yang and J.H.H.~Perk, \emph{{Eigenvectors in the superintegrable model I: generators}}, \href{https://doi.org/10.1088/1751-8113/41/27/275201}{\emph{Journal of Physics A: Mathematical and Theoretical} {\bfseries 41} (2008) 275201}.

\bibitem{AuYang09}
H.~Au-Yang and J.H.H.~Perk, \emph{{Eigenvectors in the superintegrable model II: ground-state sector}}, \href{https://doi.org/10.1088/1751-8113/42/37/375208}{\emph{Journal of Physics A: Mathematical and Theoretical} {\bfseries 42} (2009) 375208}.

\bibitem{Albertini89}
G.~Albertini, B.M.~McCoy and J.H.H.~Perk, \emph{{Level Crossing Transitions and the Massless Phases of the Superintegrable Chiral Potts Chain}}, \href{https://doi.org/10.1016/0375-9601(89)90142-4}{\emph{Phys. Lett. A} {\bfseries 139} (1989) 204}.

\bibitem{Verresen18}
R.~Verresen, N.G.~Jones and F.~Pollmann, \emph{Topology and edge modes in quantum critical chains}, \href{https://doi.org/10.1103/physrevlett.120.057001}{\emph{Physical Review Letters} {\bfseries 120} (2018) }.

\bibitem{Jones19}
N.G.~Jones and R.~Verresen, \emph{Asymptotic correlations in gapped and critical topological phases of 1d quantum systems}, \href{https://doi.org/10.1007/s10955-019-02257-9}{\emph{Journal of Statistical Physics} {\bfseries 175} (2019) 1164–1213}.

\bibitem{Albertini89c}
G.~Albertini, B.M.~McCoy, J.H.H.~Perk and S.~Tang, \emph{{Excitation spectrum and order parameter for the integrable N-state chiral Potts model}}, \href{https://doi.org/10.1016/0550-3213(89)90415-X}{\emph{Nuclear Physics B} {\bfseries 314} (1989) 741}.

\bibitem{vonGehlen96}
G.~von Gehlen, \emph{{Integrable $Z_n$-Chiral Potts Model: The Missing Rapidity-Momentum Relation}},  \href{https://arxiv.org/abs/hep-th/9601001}{{\ttfamily hep-th/9601001}}.

\bibitem{Suzuki71}
M.~Suzuki, \emph{{Relationship among Exactly Soluble Models of Critical Phenomena. I*) 2D Ising Model, Dimer Problem and the Generalized XY-Model}}, \href{https://doi.org/10.1143/PTP.46.1337}{\emph{Progress of Theoretical Physics} {\bfseries 46} (1971) 1337}.

\bibitem{Keating04}
J.~Keating and F.~Mezzadri, \emph{Random matrix theory and entanglement in quantum spin chains}, \href{https://doi.org/10.1007/s00220-004-1188-2}{\emph{Communications in Mathematical Physics} {\bfseries 252} (2004) 543–579}.

\bibitem{Smacchia11}
P.~Smacchia, L.~Amico, P.~Facchi, R.~Fazio, G.~Florio, S.~Pascazio et~al., \emph{{Statistical mechanics of the cluster Ising model}}, \href{https://doi.org/10.1103/PhysRevA.84.022304}{\emph{Phys. Rev. A} {\bfseries 84} (2011) 022304}.

\bibitem{deGottardi13}
W.~{DeGottardi}, M.~{Thakurathi}, S.~{Vishveshwara} and D.~{Sen}, \emph{{Majorana fermions in superconducting wires: Effects of long-range hopping, broken time-reversal symmetry, and potential landscapes}}, \href{https://doi.org/10.1103/PhysRevB.88.165111}{\emph{Phys. Rev. B} {\bfseries 88} (2013) 165111}.

\bibitem{Ohta16}
T.~Ohta, S.~Tanaka, I.~Danshita and K.~Totsuka, \emph{Topological and dynamical properties of a generalized cluster model in one dimension}, \href{https://doi.org/10.1103/PhysRevB.93.165423}{\emph{Phys. Rev. B} {\bfseries 93} (2016) 165423}.

\bibitem{Verresen17}
R.~Verresen, R.~Moessner and F.~Pollmann, \emph{One-dimensional symmetry protected topological phases and their transitions}, \href{https://doi.org/10.1103/PhysRevB.96.165124}{\emph{Phys. Rev. B} {\bfseries 96} (2017) 165124}.

\bibitem{Haegeman12}
J.~Haegeman, B.~Pirvu, D.J.~Weir, J.I.~Cirac, T.J.~Osborne, H.~Verschelde et~al., \emph{Variational matrix product ansatz for dispersion relations}, \href{https://doi.org/10.1103/PhysRevB.85.100408}{\emph{Phys. Rev. B} {\bfseries 85} (2012) 100408}.

\bibitem{Haegeman13}
J.~Haegeman, S.~Michalakis, B.~Nachtergaele, T.J.~Osborne, N.~Schuch and F.~Verstraete, \emph{Elementary excitations in gapped quantum spin systems}, \href{https://doi.org/10.1103/PhysRevLett.111.080401}{\emph{Phys. Rev. Lett.} {\bfseries 111} (2013) 080401}.

\bibitem{Baxter06}
R.~Baxter, \emph{{The challenge of the chiral Potts model}},  in \emph{Journal of Physics: Conference Series}, vol.~42, p.~11, IOP Publishing, 2006.

\bibitem{Baxter05}
R.J.~Baxter, \emph{{The Order Parameter of the Chiral Potts Model}}, \href{https://doi.org/10.1007/s10955-005-5534-3}{\emph{Journal of Statistical Physics} {\bfseries 120} (2005) 1–36}.

\bibitem{Tantivasadakarn23}
N.~Tantivasadakarn, R.~Thorngren, A.~Vishwanath and R.~Verresen, \emph{{Pivot Hamiltonians as generators of symmetry and entanglement}}, \href{https://doi.org/10.21468/SciPostPhys.14.2.012}{\emph{SciPost Phys.} {\bfseries 14} (2023) 012}.

\bibitem{Kramers41}
H.A.~Kramers and G.H.~Wannier, \emph{{Statistics of the Two-Dimensional Ferromagnet. Part II}}, \href{https://doi.org/10.1103/PhysRev.60.263}{\emph{Phys. Rev.} {\bfseries 60} (1941) 263}.

\bibitem{Aasen20}
D.~Aasen, P.~Fendley and R.S.K.~Mong, \emph{Topological defects on the lattice: Dualities and degeneracies},  \href{https://arxiv.org/abs/2008.08598}{{\ttfamily 2008.08598}}.

\bibitem{Fjelstad11}
J.~Fjelstad and T.~Mansson, \emph{{New symmetries of the chiral Potts model}}, \href{https://doi.org/10.1088/1751-8113/45/15/155208}{\emph{J. Phys. A} {\bfseries 45} (2012) 155208} [\href{https://arxiv.org/abs/1109.6503}{{\ttfamily 1109.6503}}].

\bibitem{Minami21}
K.~Minami, \emph{{Onsager algebra and algebraic generalization of Jordan-Wigner transformation}}, \href{https://doi.org/10.1016/j.nuclphysb.2021.115599}{\emph{Nucl. Phys. B} {\bfseries 973} (2021) 115599} [\href{https://arxiv.org/abs/2108.03811}{{\ttfamily 2108.03811}}].

\bibitem{Miao22}
Y.~Miao, \emph{{Generalised Onsager Algebra in Quantum Lattice Models}}, \href{https://doi.org/10.21468/SciPostPhys.13.3.070}{\emph{SciPost Phys.} {\bfseries 13} (2022) 070}.

\bibitem{Perk89}
J.H.H.~Perk, \emph{{Star-triangle equations, quantum Lax pairs, and higher genus curves}}, {\emph{Theta Functions--Bowdoin 1987} {\bfseries 49} (1989) 341}.

\bibitem{Davies91}
B.~Davies, \emph{{Onsager’s algebra and the Dolan--Grady condition in the non-self-dual case}}, \href{https://doi.org/10.1063/1.529036}{\emph{Journal of Mathematical Physics} {\bfseries 32} (1991) 2945}.

\bibitem{Perk16}
J.H.H.~Perk, \emph{{The early history of the integrable chiral Potts model and the odd–even problem}}, \href{https://doi.org/10.1088/1751-8113/49/15/153001}{\emph{Journal of Physics A: Mathematical and Theoretical} {\bfseries 49} (2016) 153001}.

\bibitem{Naudts09}
J.~Naudts, T.~Verhulst and B.~Anthonis, \emph{{Counting operator analysis of the discrete spectrum of some model Hamiltonians}}, \href{https://doi.org/10.1016/j.physleta.2009.07.060}{\emph{Physics Letters A} {\bfseries 373} (2009) 3419–3422}.

\bibitem{OBrien20}
E.~O'Brien, E.~Vernier and P.~Fendley, \emph{{``Not-$A$'', representation symmetry-protected topological, and Potts phases in an ${S}_{3}$-invariant chain}}, \href{https://doi.org/10.1103/PhysRevB.101.235108}{\emph{Phys. Rev. B} {\bfseries 101} (2020) 235108}.

\bibitem{Verresen25}
{R. Verresen et al}, to appear.

\bibitem{Schur18}
J.~Schur, \emph{{Über Potenzreihen, die im Innern des Einheitskreises beschränkt sind.}}, {\emph{{Journal für die reine und angewandte Mathematik}} {\bfseries 148} (1918) 122}.

\bibitem{Cohn22}
A.~Cohn, \emph{Über die anzahl der wurzeln einer algebraischen gleichung in einem kreise}, {\emph{{Mathematische Zeitschrift}} {\bfseries 14} (1922) 110}.

\bibitem{Bratelli78}
O.~Bratteli, A.~Kishimoto and D.W.~Robinson, \emph{Ground states of quantum spin systems}, \href{https://doi.org/10.1007/BF01940760}{\emph{Communications in Mathematical Physics} {\bfseries 64} (1978) 41}.

\bibitem{Fradkin80}
E.H.~Fradkin and L.P.~Kadanoff, \emph{{Disorder Variables and Parafermions in Two-dimensional Statistical Mechanics}}, \href{https://doi.org/10.1016/0550-3213(80)90472-1}{\emph{Nucl. Phys. B} {\bfseries 170} (1980) 1}.

\bibitem{Fendley12}
P.~Fendley, \emph{{Parafermionic edge zero modes in $Z_n$-invariant spin chains}}, \href{https://doi.org/10.1088/1742-5468/2012/11/p11020}{\emph{Journal of Statistical Mechanics: Theory and Experiment} {\bfseries 2012} (2012) P11020}.

\bibitem{Xu17}
W.-T.~Xu and G.-M.~Zhang, \emph{Matrix product states for topological phases with parafermions}, \href{https://doi.org/10.1103/PhysRevB.95.195122}{\emph{Phys. Rev. B} {\bfseries 95} (2017) 195122}.

\bibitem{Davis79}
P.J.~Davis, \emph{{Circulant Matrices}}, Wiley (1979).

\bibitem{Roan91}
S.-S.~Roan, \emph{{Onsager's Algebra, Loop Algebra and Chiral Potts Model}}, {\emph{MPIM Preprint Series 1991 (70).} (1991) }.

\bibitem{Nishino06}
A.~Nishino and T.~Deguchi, \emph{{The symmetry of the Bazhanov–Stroganov model associated with the superintegrable chiral Potts model}}, \href{https://doi.org/10.1016/j.physleta.2006.03.058}{\emph{Physics Letters A} {\bfseries 356} (2006) 366–370}.

\bibitem{Cardy93}
J.L.~Cardy, \emph{Critical exponents of the chiral potts model from conformal field theory}, \href{https://doi.org/10.1016/0550-3213(93)90353-Q}{\emph{Nuclear Physics B} {\bfseries 389} (1993) 577}.

\bibitem{Kato66}
T.~Kato, \emph{Perturbation theory in a finite-dimensional-space},  in \emph{{Perturbation Theory for Linear Operators}}, pp.~62--126, Springer Berlin, Heidelberg, 1966.

\bibitem{Wouters21}
J.~Wouters, H.~Katsura and D.~Schuricht, \emph{{Interrelations among frustration-free models via Witten's conjugation}}, \href{https://doi.org/10.21468/SciPostPhysCore.4.4.027}{\emph{SciPost Phys. Core} {\bfseries 4} (2021) 027}.

\bibitem{Baxter10}
R.J.~Baxter, \emph{Proof of the determinantal form of the spontaneous magnetization of the superintegrable chiral potts model}, \href{https://doi.org/10.1017/S1446181110000787}{\emph{The ANZIAM Journal} {\bfseries 51} (2010) 309–316}.

\bibitem{Alcaraz03}
F.C.~Alcaraz and M.J.~Lazo, \emph{{The Bethe ansatz as a matrix product ansatz}}, \href{https://doi.org/10.1088/0305-4470/37/1/L01}{\emph{Journal of Physics A: Mathematical and General} {\bfseries 37} (2003) L1}.

\bibitem{Katsura10}
H.~Katsura and I.~Maruyama, \emph{{Derivation of the matrix product ansatz for the Heisenberg chain from the algebraic Bethe ansatz}}, \href{https://doi.org/10.1088/1751-8113/43/17/175003}{\emph{Journal of Physics A: Mathematical and Theoretical} {\bfseries 43} (2010) 175003}.

\bibitem{Murg12}
V.~Murg, V.E.~Korepin and F.~Verstraete, \emph{{Algebraic Bethe ansatz and tensor networks}}, \href{https://doi.org/10.1103/physrevb.86.045125}{\emph{Physical Review B} {\bfseries 86} (2012) }.

\bibitem{Ruiz24}
R.~Ruiz, A.~Sopena, M.H.~Gordon, G.~Sierra and E.~L{\'o}pez, \emph{{The Bethe ansatz as a quantum circuit}}, \href{https://doi.org/10.22331/q-2024-05-23-1356}{\emph{Quantum} {\bfseries 8} (2024) 1356}.

\bibitem{Ruiz25}
R.~Ruiz, A.~Sopena, E.~López, G.~Sierra and B.~Pozsgay, \emph{{Bethe Ansatz, quantum circuits, and the F-basis}}, \href{https://doi.org/10.21468/SciPostPhys.18.6.187}{\emph{SciPost Phys.} {\bfseries 18} (2025) 187}.

\bibitem{Shibata20}
N.~Shibata, N.~Yoshioka and H.~Katsura, \emph{Onsager's scars in disordered spin chains}, \href{https://doi.org/10.1103/PhysRevLett.124.180604}{\emph{Phys. Rev. Lett.} {\bfseries 124} (2020) 180604}.

\end{thebibliography}\endgroup

\appendix

\section{Properties of \texorpdfstring{$(A_n+G_n)$}{A_n+G_n}}
\label{appendix:AplusGpsi}

To prove \cref{eqn:An+Gnpsi}, we note that the Onsager algebra admits the ladder operators $R_{m,l}^{}$ and $R_{m,l}^\dag$ given by \cite{Naudts09}

\begin{align} 
    R_{m,l} = \frac{1}{2}\big[[A_m,A_l], A_l\big] + \frac{1}{2}[A_m, A_l] = \frac{1}{4}A_m + \frac{1}{2}G_{m-l}-\frac{1}{4}A_{2l-m}\ .
\end{align}

\noindent Given an eigenstate $\ket{\psi}$ of $A_l$ with energy $E$, these obey 

\begin{align} 
    A_l (R_{m,l}\ket{\psi}) = (E -1)(R_{m,l}\ket{\psi}) \quad\quad\quad A_l (R^\dag_{m,l}\ket{\psi}) = (E +1)(R^\dag_{m,l}\ket{\psi})\ .
\end{align}

\noindent It follows that 

\begin{align}
\begin{split}
    A_0 (A_n+G_n)\ket{\psi_0} &=(A_n+G_n)A_0\ket{\psi_0} + [A_0, A_n+G_n] \ket{\psi_0}\\
    &= E_0 (A_n+G_n)\ket{\psi_0} -2 R_{n, 0} \ket{\psi_0}\ .
\end{split}
\end{align}

\noindent However, since $\ket{\psi_0}$ is the ground state of $A_0$, there exists no eigenstate with energy $E_0-1$. Therefore, $R_{n,0} \ket{\psi_0} = 0$, and we have 

\begin{align}
    A_0 (A_n+G_n)\ket{\psi_0} = E_0 (A_n+G_n)\ket{\psi_0}\ .
\end{align}

\noindent Thus, acting with $(A_n+G_n)$ on $\ket{\psi_0}$ keeps us within the ground state sector of $A_0$. Given that this ground state is unique, we must have

\begin{align}
    (A_n+G_n)\ket{\psi_0} \propto \ket{\psi_0}\ .
\end{align}

Additionally, any eigenstate $\ket{\psi}$ of $A_0$ with energy $E$ that is annihilated by $R_{n,0}$ obeys

\begin{align}
    A_0 (A_n+G_n)\ket{\psi} = E (A_n+G_n)\ket{\psi}\ .
\end{align}

\noindent Thus, $(A_n+G_n)\ket{\psi}$ acting on $\ket{\psi}$ keeps us within the eigenspace of $A_0$ with energy $E$. Hence, if all states within this eigenspace obey $R_{n,0}\ket{\psi} = 0$, it is possible to find a basis $\{\ket{\phi_E^{(i)}}\}$ such that 

\begin{align} \label{eqn:An+Gn-excited}
    (A_n+G_n)\ket{\phi_E^{(i)}} \propto \ket{\phi_E^{(i)}}\ .
\end{align}

\noindent Given that the lowest charge $m$ excitations have energy $E_0+m/N$ \cite{Jones24}, and that $R_{n,0}\ket{\phi_E} \propto \ket{\phi_{E-1}}$, states with $m<N$ can be put in bases satisfying \eqref{eqn:An+Gn-excited}, as $R_{n,0}$ must annihilate all these states. 

We note that each $(A_n+G_n)$ commutes with the translation operator. Thus, for single-particle excitations, for which there are $L$ possible states, the appropriate basis of states $\{\ket{\phi^{(P)}_{E_0+1/N}}\}$ that satisfy \eqref{eqn:An+Gn-excited} is given by the $L$ eigenstates of the translation operator

\begin{align}
    \ket{\phi^{(P)}_{E_0+1/N}} = \frac{1}{\sqrt{L}}\sum_{j=1}^L e^{-i P j} Z_j^\dag \ket{\psi_0} \quad\quad\quad P \in \bigg\{0, \frac{2\pi}{L}, \frac{4\pi}{L} , \dots ,  \frac{2(L-1)\pi}{L}\bigg\}\ .
\end{align}

\noindent We therefore have

\begin{align}
    (A_n+G_n) \ket{\phi^{(P)}_{E_0+1/N}} \propto \ket{\phi^{(P)}_{E_0+1/N}}\ .
\end{align}

\section{Disorder operator for \texorpdfstring{$d=1$}{d=1}}
\label{appendix:string-order-details}

For even $N$, the unitary disorder operator is given by

\begin{align}
    \mu_k  = \prod_{j=-\infty}^k X_j^{N/2}\ .
\end{align}

\noindent Choose $|a|<1$ so that $\beta \in \mathbb{R}$. We aim to evaluate

\begin{align}
    \braket{\mu_k} = \frac{\bra{\varphi}\mu_k\ket{\varphi}}{\braket{\varphi|\varphi}} = \frac{\bra{\psi_0}e^{-\beta A_1}\mu_ke^{-\beta A_1}\ket{\psi_0}}{\braket{\psi_0|e^{-2\beta A_1}|\psi_0}}\ .
\end{align}

\subsection{Simplifying the numerator} \label{appendix:string-order-details-numerator}

In this section, we simplify

\begin{align}
    \braket{\varphi|\mu_k|\varphi} = \braket{\psi_0|e^{-\beta A_1} \mu_ke^{-\beta A_1} |\psi_0}\ .
\end{align}

\noindent Firstly, define

\begin{align}
    A_1 = -\frac{1}{N} \sum_{j=1}^L \sum_{m=1}^{N-1} \alpha_m (Z_{j}^\dag Z_{j+1}^{})^m
\end{align}

\noindent and

\begin{align}
    U_{j,j+1}^{(m)}(\beta) = \exp\left(\frac{\beta}{N}\alpha_m(Z_{j}^\dag Z_{j+1})^m\right)\ ,
\end{align}

\noindent so that

\begin{align}
    e^{-\beta A_1} = \prod_{j=1}^L \prod_{m=1}^{N-1} U_{j, j+1}^{(m)}(\beta)\ .
\end{align}

\noindent Moreover, since $\mu_k$ is unitary, we have

\begin{align}
    \mu_k e^{-\beta A_1}  \mu_k^\dag = e^{-\beta\mu_k A_1 \mu_k^\dag}\ .
\end{align}

\noindent Given that $X_j Z_k = \omega^{\delta_{jk}}Z_k X_j$, we have 

\begin{align}
    \mu_k A_1 \mu_k^\dag = A_1  + \frac{2}{N}\sum_{\substack{m=1 \\ m \text{ odd}}}^{N-1} \alpha_m (Z_{k}^\dag Z_{k+1}^{})^m\ .
\end{align}

\noindent Therefore,

\begin{align}
    \mu_k e^{-\beta A_1}= e^{-\beta\mu_k A_1 \mu_k^\dag} \mu_k= e^{-\beta A_1} \prod_{\substack{m=1 \\ m \text{ odd}}}^{N-1} U^{(m)}_{k,k+1}(-2\beta) \mu_k\ .
\end{align}

\noindent Given that $\mu_k \ket{\psi_0} = \ket{\psi_0}$, we arrive at

\begin{align}
    \bra{\psi_0}e^{-\beta A_1}\mu_ke^{-\beta A_1}\ket{\psi_0}=\bra{\psi_0}e^{-2\beta A_1} \prod_{\substack{m=1 \\ m \text{ odd}}}^{N-1}U_{k,k+1}^{(m)}(-2\beta) \ket{\psi_0}\ .
\end{align}

\subsection{The transfer matrix method}

\noindent If we write $\ket{\psi_0}$ in terms of the states $\ket{a}$ that are diagonal in the $Z$-basis

\begin{align}
    \ket{\psi_0} = \frac{1}{\sqrt{N^L}}\sum_{a_1\dots a_L =0}^{N-1} \ket{a_1 \dots a_L}\ ,
\end{align}

\noindent then

\begin{align}
    {\braket{\psi_0|e^{-2\beta A_1}|\psi_0}} = \frac{1}{N^L}\sum_{a_1 \dots a_L=0}^{N-1}\prod_{j=1}^L\exp\left(\frac{2\beta }{N} F_1(a_{j-1}, a_j)\right)
\end{align}

\noindent and 

\begin{align}
    \bra{\psi_0}e^{-2\beta A_1} \prod_{\substack{m=1 \\ m \text{ odd}}}^{N-1}U_{k,k+1}^{(m)}(-2\beta) \ket{\psi_0} =\frac{1}{N^L}\sum_{a_1 \dots a_L=0}^{N-1}\prod_{j=1}^L\exp\left(\frac{2\beta }{N} F_1(a_{j-1}, a_j)\right) \exp \left(-\frac{2\beta}{N} F_2(a_k, a_{k+1})\right)\ ,
\end{align}

\noindent where we have defined 

\begin{align}
    F_1(x,y) = \sum_{m=1}^{N-1}\alpha_m \omega^{-m(x-y)} = \begin{cases}
                \frac{N-1}{2}-(x-y) &  0 \leq x-y < N \\
                \frac{-N-1}{2}-(x-y) &  -N \leq x-y < 0
            \end{cases}
\end{align}

\begin{align}
    F_2(x,y) = \sum_{\substack{m=1 \\ m \text{ odd}}}^{N-1}\alpha_m \omega^{-m(x-y)} = \begin{cases}
            N/4 & 
            \begin{array}[c]{l}
            0 \leq x - y < N/2 \\ -N \leq x - y < -N/2
            \end{array} \\
            -N/4 & 
            \begin{array}[c]{l}
            -N/2 \leq x - y < 0 \\
             N/2 \leq x - y < N\ .
            \end{array}
    \end{cases}
\end{align}

\noindent The simplification for $F_1(x,y)$ follows from a trigonometric identity, given as Eq. (27) in Ref. \cite{Jones24}. The simplification for $F_2(x,y)$ follows from writing 

\begin{align}
    F_2(x,y) = F_1(x,y) - \sum_{\substack{m=1 \\ m \text{ even}}}^{N-1}\alpha_m \omega^{-m(x-y)}\ .
\end{align}

\noindent The second sum can be rearranged into a form analogous to $F_1(x,y)$, which can be similarly simplified with trigonometric identities.

If we define the matrices with entries

\begin{align}
    T_{a,b} = \exp\left(\frac{2\beta}{N} F_1(a,b)\right) \quad\quad\quad B_{a,b} = \exp\left(\frac{2\beta}{N} \left(F_1(a,b)-F_2(a,b)\right)\right)\ ,
\end{align}

\noindent then, using the cyclic property of the trace,

\begin{align}
    \braket{\mu_k} = \frac{\text{Tr}(T^{L-1} B)}{\text{Tr}(T^L)} \implies \lim_{L\rightarrow \infty}\braket{\mu_k} = \frac{\vec{v}_0^T B \vec{v}_0}{\lambda_0}\ ,
\end{align}

\noindent since, in the thermodynamic limit $L \rightarrow \infty$, the largest eigenvalue $\lambda_0$ of the transfer matrix $T$, with eigenvector $\vec{v}_0$, dominates. 

For general even $N$, we find the circulant transfer matrix

\begin{align}
    T = \begin{pmatrix}
            e^{\beta (N-1)/N} & e^{-\beta (N-1)/N} & \cdots \\
            e^{\beta (N-3)/N} & e^{\beta (N-1)/N} & \cdots \\
            e^{\beta (N-5)/N} & e^{\beta (N-3)/N} & \cdots \\
            \vdots & \vdots & \ddots & \\
            e^{-\beta (N-1)/N} & e^{-\beta (N-3)/2N} & \cdots \\
        \end{pmatrix}.
\end{align}

\noindent Each column in this matrix is a cyclic permutation of the first one, which has $N$ entries $T_k$ defined by

\begin{equation}
    T_k = e^{\beta(N-(2k+1))/N} \quad\quad\quad \text{for} \quad\quad\quad k \in \{0,1,\dots,N-1\}\ .
\end{equation}

\noindent The matrix $B$ is similarly circulant, with each column being a cyclic permutation of the initial column, having $N$ entries $B_k$ defined by

\begin{equation}
    B_k = \begin{cases}
                e^{-\beta/2}T_k \quad &\text{for} \quad k \in \{0,1,\dots,N/2-1\}\\
                e^{\beta/2}T_k \quad &\text{for} \quad k \in \{N/2,\dots,N-1\}
    \end{cases}\ .
\end{equation}

\noindent A circulant matrix can be diagonalised using standard results \cite{Davis79}. By the triangle inequality, the largest eigenvalue of $T$ (and $B$) corresponds to the eigenvector

\begin{align}
    \vec{v}_0 = \frac{1}{\sqrt{N}}(1,1,\dots,1)\ .
\end{align}

\noindent Thus, using $\beta=2 \,\text{arctanh}(a)$, we obtain

\begin{align}
\begin{split}
    \lim_{L\to\infty}\braket{\mu_k} &= \frac{\sum_{k=0}^{N/2-1} \cosh \left(-\frac{\beta}{2N}(N-2(2k+1))\right)}{\sum_{k=0}^{N/2-1}\cosh \left(-\frac{\beta}{2N}(2N-2(2k+1))\right)}\\
    &= \frac{1}{\cosh(-\beta/2)}\\
    &= \sqrt{1-a^2}
\end{split}
\end{align}

\noindent for all even $N$.

\section{Algorithm for finding the coefficients \texorpdfstring{$\{r_\alpha\}$}{r_α}}
\label{appendix:algorithm-t}

Should we wish to calculate the coefficients in \cref{eqn:Htilde-Majoranas}, this can be done algorithmically as shown below in Algorithm \ref{alg:compute-talpha}. To demonstrate the operation of this algorithm, we use \eqref{eqn:layer-operation} to find

\begin{align}
\begin{split}
    \mathrm{M}^{(d)^{-1}}{\Omega}_n \mathrm{M}^{(d)} &\propto \sum_{\alpha=0}^d s_\alpha\left(\gamma_{n+\alpha}+i \tilde{\gamma}_{n-\alpha}\right) + \sum_{\alpha=0}^d b_d s_{d-\alpha}\left(\gamma_{n+\alpha}+i \tilde{\gamma}_{n-\alpha}\right)\\
    &\propto \sum_{\alpha=0}^d r_\alpha^{(d)}\left(\gamma_{n+\alpha}+i \tilde{\gamma}_{n-\alpha}\right)\ .
\end{split}
\end{align}

\noindent The next layer yields

\begin{align}
\begin{split}
    \mathrm{M}^{(d-1)^{-1}}\mathrm{M}^{(d)^{-1}}{\Omega}_n \mathrm{M}^{(d)}\mathrm{M}^{(d-1)}
    &\propto \sum_{\alpha=0}^d r_\alpha^{(d)}\left(\gamma_{n+\alpha}+i \tilde{\gamma}_{n-\alpha}\right)
    +b_{d-1}\sum_{\alpha=-1}^{d-1}r_{d-1-\alpha}^{(d)}\left(\gamma_{n+\alpha}+i \tilde{\gamma}_{n-\alpha}\right)\\
    &\propto \sum_{\alpha = -1}^d r_\alpha^{(d-1)}\left(\gamma_{n+\alpha}+i \tilde{\gamma}_{n-\alpha}\right)\ ,
\end{split}
\end{align}

\noindent with the third layer giving

\begin{align}
\begin{split}
    \mathrm{M}^{(d-2)^{-1}}\mathrm{M}^{(d-1)^{-1}}\mathrm{M}^{(d)^{-1}}{\Omega}_n \mathrm{M}^{(d)}\mathrm{M}^{(d-1)}\mathrm{M}^{(d-2)}
    &\propto \sum_{\alpha=-1}^d r_\alpha^{(d-1)}\left(\gamma_{n+\alpha}+i \tilde{\gamma}_{n-\alpha}\right)\\
    &\quad +b_{d-2}\sum_{\alpha=-2}^{d-1}r_{d-2-\alpha}^{(d-1)}\left(\gamma_{n+\alpha}+i \tilde{\gamma}_{n-\alpha}\right)\\
    &\propto \sum_{\alpha = -2}^d r_\alpha^{(d-2)}\left(\gamma_{n+\alpha}+i \tilde{\gamma}_{n-\alpha}\right)\ ,
\end{split}
\end{align}

\noindent where we have defined $r_\alpha^{(k)} = r_\alpha^{(k+1)} + b_k r_{k-\alpha}^{(k+1)}$ and $r_\alpha^{(d+1)} = s_\alpha$. Proceeding in this manner gives Algorithm \ref{alg:compute-talpha}. Note that the second sum initially has an upper limit of $d$, but for every following layer, it has an upper limit of $d-1$. This follows directly from Eq. (\ref{eqn:layer-operation}).

\begin{algorithm}[H]
\caption{Compute coefficients $\{r_\alpha\}$ from vector $\vec{s} = (s_0, \dots, s_d)$}\label{alg:compute-talpha}
\begin{algorithmic}
\Require Vector $\vec{s} = (s_0, \dots, s_d)$ with $s_0 \neq 0$
\Require Vector \texttt{bList} $= (b_1, \dots, b_d)$ with all $|b_k| \neq 1$  
from Algorithm \ref{alg:compute-bk} 
\Ensure Returns a list proportional to $(r_{1-d}, \dots, r_d)$

\State Initialize list \texttt{rList} $ \gets (r_{1-d}, \dots, r_{d})$ where $(r_0, \dots, r_d) \gets \vec{s}$ and all other $r_\alpha \gets 0$
\For{$k = d, d-1, \dots, 1$} 
    \State Initialise \texttt{uList} $= (u_{1-d},\dots,u_d) \gets$ \texttt{rList}
    \If{$k = d$}
        \State $i_\text{max} \gets d$
    \ElsIf{$k<d$}
        \State $i_\text{max} \gets d-1$
    \EndIf
    \For{$i = k-d, \dots, i_\text{max}$}
        \State $u_i \gets r_i + b_k \cdot r_{k - i}$
    \EndFor
    \State \texttt{rList} $\gets$ \texttt{uList}
\EndFor
\State \Return \texttt{rList} \Comment{This output is proportional to the vector of coefficients $\vec{r} = (r_{1-d},\dots, r_d)$}
\end{algorithmic}
\end{algorithm}
\section{Understanding odd \texorpdfstring{$p$}{p}}
\label{appendix:kramers-wannier}

\subsection{Handling degeneracy}

Consider generators $\{A_l, G_m|l,m\in\mathbb{Z}\}$ obeying the Onsager algebra. As noted in Section \ref{sec:results}, making the replacements $A_k \rightarrow A_{k+1}, G_k \rightarrow G_k$ gives a new set of generators obeying the Onsager algebra. As our proof in Section \ref{subsec:theorem-1-proof} relies only on the structure of the Onsager algebra and the uniqueness of the ground state, given that 

\begin{align}
    H_A^{p=0} = \sum_{k,q=0}^d s_k s_q A_{k+q}
\end{align}

\noindent has an eigenstate $\ket{\varphi^{p=0}} = e^{-\beta_d A_d}\dots e^{-\beta_1 A_1} \ket{\psi_0}$, applying the above mapping between the generators $p$ times, we can conclude that 

\begin{align}
    H_A^{p} = \sum_{k,q=0}^d s_k s_q A_{k+q+p}
\end{align}

\noindent has an eigenstate 

\begin{align}
    \ket{\varphi^{p}} = e^{-\beta_d A_{d+p}}\dots e^{-\beta_1 A_{1+p}} \ket{\psi_{p}}.
\end{align}

This is of the form given in \cref{result:theorem-1}. However, we have ignored a subtlety: for $\ket{\psi_0}$ to be an eigenstate of each $(A_{\alpha} + G_\alpha)$, we used the fact that the ground state is unique. This is discussed in Appendix \ref{appendix:AplusGpsi}. As $A_{2k}$ is related to $A_{0}$ by the unitary transformation \eqref{eqn:unitary-pivot}, the ground state is also unique for even $p$, and the argument above holds. 

However, for clock models, $A_1$ instead has a set of $N$ degenerate ground states $\{\ket{\psi_1^{(n)}}\}$; any $A_{2k+1}$, being unitarily related to $A_1$, will similarly have ground-state degeneracy $N$. Therefore, our argument breaks down unless we can find a basis of $\{\ket{\psi_1^{(n)}}\}$ that simultaneously diagonalises every $(A_{\alpha+1}+G_\alpha)$, as we now have (following the argument in Appendix \ref{appendix:AplusGpsi})

\begin{align}
    (A_{\alpha+1} + G_\alpha)\{\ket{\psi_1^{(n)}}\} = \sum_m c_m \{\ket{\psi_1^{(m)}}\}
\end{align}

\noindent for some constants $\{c_m\}$. However, as noted in Section \ref{sec:spectrum}, the operator   $r= \prod_{j=1}^L X_j$
\noindent commutes with every generator $A_l, G_m$. Thus, for all $\alpha$, the appropriate basis diagonalising $(A_{\alpha+1}+G_\alpha)$ must also diagonalise $r$. 

Working in the $Z$-diagonal basis, $A_1$ has $N$ degenerate ground states $\ket{\tau_a} = \ket{aaa\dots}$ for $a \in \{0,1,\dots,N-1\}$. The unique basis that diagonalises $r$ is

\begin{align}
    \ket{\psi_1^{(n)}} = \sum_{a=0}^{N-1} \omega^{-na} \ket{\tau_a} \quad\quad\quad n\in\{0,1,\dots,N-1\}\ .
\end{align}

\noindent Each of these states has charge $Q=n$. As $r$ commutes with all $A_l, G_m$, the basis $\{\ket{\psi_1^{(n)}}\}$ with distinct charges must diagonalise $(A_{\alpha+1}+G_\alpha)$ for all $\alpha$. Thus, despite the fact that $A_1$ does not have a unique ground state, \cref{result:theorem-1} still holds. The pivot procedure \eqref{eqn:unitary-pivot} allows us to draw the same conclusion for all odd $p$.

\subsection{A note on chain length}

If we wish to find the ground state of a Hamiltonian with $f(z) = -z^p g(z)^2$, we can perform a similar analysis by replacing $A_k \rightarrow -A_k$ and $\ket{\psi_p} \rightarrow \ket{\psi_p^-}$. For odd $p$, we again encounter the obstacle of degeneracy. For $-A_1$, if $L = 0 \mod N$, we have $N$ degenerate ground states of the form $\ket{\upsilon} = T^{\upsilon}\ket{0,1,2,\dots,N-1,0,1,2,\dots}$, where $T$ is the translation operator and $\upsilon \in \{0,1,\dots,N-1\}$. As before, we can put these states into a basis that simultaneously diagonalises every $A_{\alpha+1}+G_\alpha$:

\begin{align}
    \ket{\psi_1^{(n),-}} = \sum_{\upsilon=0}^{N-1} \omega^{n\upsilon}\ket{\upsilon}\ .
\end{align}

However, if we do not have $L = 0 \mod N$, we obtain frustrated ground states, making a similar analysis more difficult. We choose $L=0\mod N$ to avoid this complication.
\section{Zeros on the unit circle}
\label{appendix:schur-cohn-algorithm}

In this appendix, we make the connection between zeros of $g(z)$ on the unit circle, and having some $|b_k|=1$ appearing when we apply Algorithm \ref{alg:compute-bk}.

We begin with the polynomial $g(z)$ of degree $d$ and its reciprocal polynomial $\tilde{g}(z) = z^d g(1/z)$ given by

\begin{align}
    g(z) = \sum_{k=0}^d s_k z^k \quad\quad\quad \tilde{g}(z) = \sum_{k=0}^d s_{d-k}z^k\ ,
\end{align}

\noindent where the coefficients $s_k$ are real. Assume further that $s_0, s_d\neq0$. It follows that $g(z)$ and $\tilde{g}(z)$ must share any zeros on the unit circle (along with their multiplicities).

Now define the Schur transform, which is a polynomial of the form

\begin{align}
\begin{split}
    Tg(z) &= g(z) - \frac{\tilde{g}(0)}{g(0)}\tilde{g}(z)\\ 
    &= \sum_{k=0}^d \left(s_k - \frac{s_d}{s_0} s_{d-k}\right) z^k\\
    &= \sum_{k=0}^{d-1} s^{(d)}_k z^k,
\end{split}
\end{align}

\noindent for $s^{(d)} = (s_k - s_d s_{d-k}/s_0)$. Since $s^{(d)}_d=0$, $Tg(z)$ is a polynomial of lower degree than $g, \tilde{g}$. Defining $T^0g(z) = g(z)$, we can iteratively construct the series of polynomials for $0\leq n\leq d$

\begin{align}
    T^n g(z) = T \left(T^{n-1}g(z)\right) = \sum_{k=0}^{d-n}s_k^{(d-(n-1))}z^k\ ,
\end{align}

\noindent where 

\begin{align}
    s_k^{(m)} = s_k^{(m+1)} - \frac{s_{m}^{(m+1)}}{s_0^{(m+1)}} s_{m-k}^{(m+1)}\quad\quad\quad \text{and}\quad\quad\quad s^{(d+1)}_k = s_k\ .
\end{align}

\noindent Moreover, Algorithm \ref{alg:compute-bk} defines

\begin{align}
    b_m = \frac{s_m^{(m+1)}}{s_0^{(m+1)}}\ .
\end{align}

\noindent In general, each polynomial $T^{n-1}g(z)$ has lower degree than the polynomial $T^ng(z)$.

From these definitions, it follows that $g(z)$ and $\tilde{g}(z)$ having a root $z_1=e^{i\theta}$ on the unit circle implies either $Tg(z_1) = \widetilde{Tg}(z_1) = 0$ or $g(0)=0$ (giving undefined $Tg(z)$). By iteration, for all $n$, either $T^ng(z_1) = 0$ or $T^{n-1}g(0) = 0$ so that $T^ng(z)$ is undefined. 

Firstly, consider the case where $T^ng(z)$ is first undefined for some $n=p+1$, where $0<p \leq d-1$.\footnote{Note that we have defined $g(z)$ from $f(z)$ such that the zeroth order term in $g(z)$ does not vanish. Therefore, $Tg(z)$ is always defined.} This imposes that 

\begin{align}
    T^{p}g(0)=s_0^{(d+1-p)}=0\implies
    \left(s_0^{(d+2-p)}\right)^2=\left(s_{d+1-p}^{(d+2-p)}\right)^2\ .
\end{align}

\noindent Thus, 

\begin{align}
    |b_{d+1-p}|= \left|\frac{s_{d+1-p}^{(d+2-p)}}{s_0^{(d+2-p)}}\right| = 1\ ,
\end{align}

\noindent and so any undefined $T^ng(z_1)$ for $0< n\leq d$ will give at least one $|b_m|=1$ for $2\leq m\leq d$.

Now consider the alternate case where $T^ng(z)$ is defined for all $0\leq n\leq d$. If $z_1 = e^{i\theta}$ is a root of $g(z)$, then it is a root of all $T^ng(z)$ for all $n\leq d$. Thus, the final degree-zero polynomial must satisfy

\begin{align}
    T^d(z_1) = s_0^{(1)} = 0 \implies \left(s_0^{(2)}\right)^2=\left(s_{1}^{(2)}\right)^2\ .
\end{align}

\noindent This gives 

\begin{align}
    |b_1| = \left|\frac{s_{1}^{(2)}}{s_0^{(2)}}\right| = 1\ .
\end{align}

\noindent Therefore, if $g(z)$ has a root $z_1 = e^{i\theta}$ on the unit circle, Algorithm \ref{alg:compute-bk} must give $|b_m|=1$ for some $1\leq m \leq d$. As we wish to exclude cases where this occurs, we will only consider polynomials $g(z)$ without roots on the unit circle. Note, however, that the converse does not hold; having some $|b_k|=1$ does not imply that $g(z)$ has a root on the unit circle. For example, consider $g(z) = 1-4z+z^2$: this gives $b_2 = 1$, but has roots $z_\pm = 2 \pm \sqrt{3}$ with $|z_\pm| \neq 1$.
\end{document}